\documentclass[onecolumn,preprintnumbers,amsmath,amssymb,floatfix]{revtex4}

\usepackage{graphicx, latexsym, float}
\usepackage{color} 
\usepackage{dcolumn}
\usepackage{bm}
\usepackage{subfigure}
\definecolor{orange}{rgb}{1,0.5,0}
\definecolor{light-gray}{gray}{0.95}

\begin{document}


\title{Structurally Robust Control of Complex Networks}

\author{Jose C. Nacher\footnotemark[2]}
 \email{nacher@is.sci.toho-u.ac.jp}
\affiliation{
Department of Information Science, Faculty of Science, \\Toho University, Miyama 2-2-1, Funabashi, Chiba 274-8510, Japan
}%

\author{Tatsuya Akutsu\footnote{JCN and TA are corresponding authors}}
\email{takutsu@kuicr.kyoto-u.ac.jp}
\affiliation{
Bioinformatics Center, Institute for Chemical Research, \\Kyoto University, Uji, 611-0011, Japan
}%

\date{\today}

\begin{abstract}
Robust control theory has been successfully applied to numerous real-world problems  
using a small set of devices called {\it controllers}. However, the real systems represented by networks contain unreliable 
components and modern robust control engineering has not addressed the problem of structural changes on a large network. Here, 
we introduce the concept of structurally robust control of complex networks and provide a concrete example using an algorithmic framework
that is widely applied in engineering. The developed analytical tools, computer simulations and real network analyses lead 
herein to the discovery 
that robust control can be achieved in scale-free networks
with exactly the same order of controllers required in a standard non-robust configuration by adjusting 
only the minimum degree. The presented methodology also addresses 
the probabilistic failure of links in real systems, such as neural synaptic unreliability in {\it C. elegans},  
and suggests a new direction to pursue in studies of complex networks in which control theory has a role.
\end{abstract} 



\maketitle



\section{Introduction}

Real networks contain unreliable components; in critical infrastructures and technological networks some links may become
non-operational owing to disasters or accidents, and in natural networks this might occur owing to pathologies.
Although the robustness and resilience of networks have been extensively investigated 
over the past decade \cite{newman, havlin1, havlin2, reka}, controllability 
methods for complex networks that can robustly manage structural changes have not been developed. The existing research is limited
to recent studies of network controllability under node attack and cascading failures using maximum matching \cite{baraliu, pu, nie},
the discussion of quantitative measures of network robustness to investigate the effect
of edge removal on the number of controllable nodes without a comprehensive theoretical analysis \cite{ruths} and studies on multi-agent
systems under simultaneous failure of links and agents \cite{rahimian}. Note that
the problem of how the number of driver nodes change as function of removal fraction of edges \cite{pu, nie, ruths} 
and our question of how to design complex networks with structurally robust control feature are drastically different. 
While the former are heavily relying 
on percolation and cascading failure techniques well-studied over the past decade, our work 
studies the minimum number of driver nodes
to control the entire network against arbitrary
single or multiple edge failures.

Here, 
we introduce the concept of structurally robust control
of complex networks. To provide a concrete example, we adopt the Minimum Dominating Set (MDS) model
because it has been widely applied to the control of engineering systems, such as mobile ad hoc 
networks (MANET), transportation 
routing, computer communication networks \cite{haynes, sto, hawaii, kumar, book1, karbasi}, 
design of swapped networks for constructing large parallel and distributed systems \cite{swapped}
as well as the investigation of social influence propagation \cite{social1, social2}. Robust 
control theory emerged in the late 1970s and is based on linear, time-invariant transfer functions. The controller 
is designed to change the system's model dynamics until it reaches a certain degree of uncertainty or disturbance. Thus, the 
system is designed to be robust or stable against the presence of bounded modelling errors. To address 
disturbances, techniques such as single-input, single-output (SISO) feedback and H-infinity loop-shaping were 
developed to avoid dynamic trajectories that deviate when disturbances enter the system \cite{robust, robust2}. Although modern
robust control engineering has been 
successfully applied to numerous real-world problems, such as stability in aircrafts and satellites and the efficiency 
of power, manufacturing, and chemical plants, it has 
not yet been applied to the problem of structural changes on a large network. 

\begin{figure}[th]
\includegraphics[angle=0, scale=0.6]{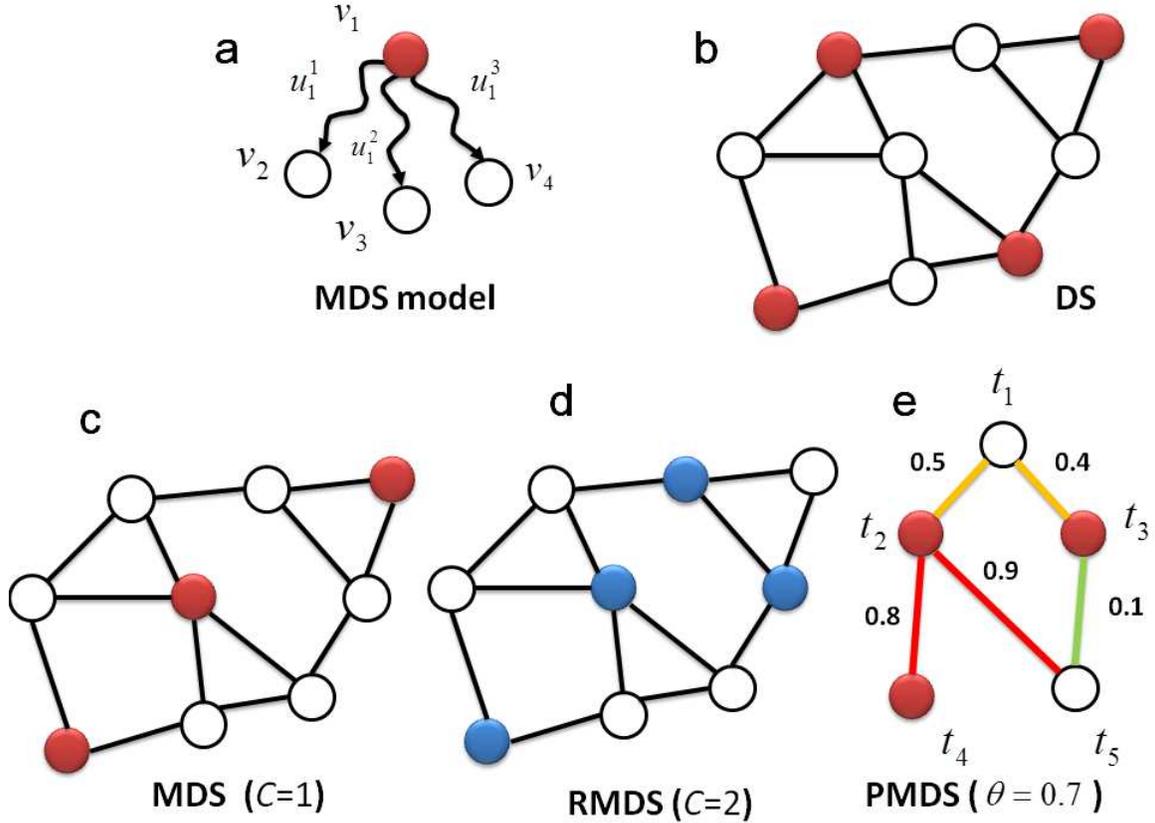}
\caption{ \label{fig:schema} {\bf The MDS model and the robust domination configuration.} (a) The network is structurally controllable 
by selecting a Minimum Dominating Set (MDS) because each
dominated node has its own control signal. A maximum matching approach needs three
driver nodes $v_1$,$v_2$, and $v_4$, assuming a matching link from $v_1$ to $v_3$. In contrast,
the MDS only requires one node, $v_1$. The labels $u_1^1$, $u_1^2$ and $u_1^3$ indicate control. (b) Example of a Dominating Set (DS) 
in a graph $G$. A set of nodes $S$ (filled) in a graph $G$ is a dominating set if every
node in $G$ is either an element of $S$ or adjacent to an element of $S$. (c) The graph $G$ allows for an MDS 
with cover $C$=1 composed of only three nodes. (d) Example of a Robust Minimum Dominating set (RMDS)
(i.e., an MDS with cover $C$=2). In an RMDS, each node in $G$ is either an element of $S$ or adjacent to at least two elements of $S$.
(e) Each edge ($t_i$, $t_j$) has a failure probability of $P_{t_i,t_j}$ highlighted in the figure. 
In a Probabilistic Minimum Dominating Set (PMDS), the approach shown here, each node must 
be covered by multiple nodes in the MDS so that the probability 
that at least one edge is active is at least $\theta=0.7$. For instance, $t_4$ must cover itself because the link ($t_2, t_4$) is
unreliable with a failure probability of 0.8, which is higher than $\theta=0.7$.}
\end{figure}

A set of nodes $S$ in a graph $G$ is a dominating set (DS) if every node in $G$ is either an element of $S$ or adjacent to 
an element of $S$. Then, the MDS approach states that a network 
is made structurally controllable by selecting an MDS (driver set) because each dominated node has its own control signal 
\cite{nacher1, nacher2, nacher3} (see Fig. ~\ref{fig:schema}a). Recently, 
Moln\'{a}r et al. further studied the size of an MDS by exhaustively comparing several types 
of artificial scale-free networks using a greedy algorithm \cite{molnar}. Interestingly, 
Wuchty demonstrated the applicability of the MDS approach \cite{nacher1} to the
 controllability of protein interaction networks and 
showed that the Minimum Dominating Set of proteins were enriched with essential, cancer-related and virus-targeted genes \cite{wuchty}.
Whereas each element is controlled by at least one node in $G$ ($C$=1) (or is covered by itself) in an MDS, 
the novel robust MDS (RMDS) approach states that each node must be covered by itself or at least two nodes 
in $G$ ($C$=2) (see Fig. ~\ref{fig:schema}cd). The analytical results and computer simulations 
demonstrate that a robust configuration ($C$=2, $D$=2) and non-robust configuration ($C$=1, $D$=1) of 
a scale-free network with minimum degree $D$ require the same order of driver nodes. The robust configuration guarantees that 
the system remains controllable 
even under arbitrary single or multiple link failure. This finding has remarkable implications
for designing technical and natural systems that can still operate in the presence of unavailable or damaged links 
because the implementation of such a robust system in a large network does not change the order of the required controllers in 
a conventional system without robustness capability. As a byproduct of this research, our results also demonstrate that the minimum 
degree $D$ in a network plays an important role in network controllability and significantly affects the size of the MDS. In particular, 
for $\gamma < 2$, the order of the size of an MDS changes if the minimum degree changes, unveiling another tool to reduce 
the number of driver nodes. These theoretical findings are confirmed by computer simulations and an analysis of real-world 
undirected, directed and bipartite networks. In addition, the MDS approach is extended to address probabilistic network domination 
when we consider the probability of link transmission failure. The derived mathematical tools allow us to identify
optimal controllability configurations in real biological systems, by mapping the synaptic unrealiability distribution
experimentally observed in rat brains \cite{failure} to the most well-known 
and recently updated neural network model for {\it C. elegans} organism \cite{neuron}. 

The concept of structural controllability was first introduced by Lin \cite{lin} for single-input systems and it was quickly
extended to multi-input systems \cite{shields, hosoe, maeda, murota}. The maximum matching (MM) algorithm identifies the minimum number of nodes
to control the entire network, by providing a mapping between structural controllability and network structure \cite{baraliu}. However, 
there are several
striking differences with MDS approach: (1) By using MM approach, the fraction of driver nodes tends to be minimized in random networks.
The MDS does not necessarily give a minimum number of driver nodes in the sense of MM approach. 
However, MDS gives less number of driver nodes 
in many cases, including scale-free networks in which hubs are present \cite{newman, caldarelli}. For example, consider the star graph 
(all nodes but one node are leaves) with $n$ leaves. Then, MM approach needs $n$-1 driver nodes whereas 
MDS approach needs only one driver node. (2) The MM approach is based on 
linear systems whereas MDS approach does not even need structural controllability: 
it is enough to assume that a node is controllable if it is directly connected to a driver node. This represents 
one of the unique features of the MDS model because it
suggests that can be applied to a certain kind of nonlinear
and/or discrete models. However, these striking advantages have a price.  
(1) The set of driver nodes in MDS is $O(n)$ in scale-free networks with $\gamma >2$. (2) Each edge has to be controlled independently. 
However, even in the case of $O(n)$, a relatively small number of drivers is required in most cases. Above all, 
the proposed concept of structurally robust control 
is an algorithmic-independent framework. Therefore, engineering applications of structurally robust control 
may flexibly give preference to one algorithm over another. 
\section{THEORETICAL RESULTS for robust domination (RMDS)}

\subsection{Analysis for the case $\gamma<2$ with minimum degree $D$ and MDS with cover $C$=1}
\label{sec:gbig}

In the following, we present analytically derived predictions for the minimum number of drivers using the 
MDS controllability approach by considering specific cases for the degree exponent $\gamma$ and the minimum degree $D$. Then, 
the robust control is also analysed by considering the number of drivers $C$ required to cover each node. We first assume 
that the minimum degree of an undirected graph $G(V,E)$ with $n$ nodes and a 
degree distribution that follows a power-law $P(k)=\alpha k^{-\gamma}$ is $D$. We then use a standard mean-field approach that assumes a continuum approximation 
for the degree $k$, so that it becomes a continuous real variable \cite{newman, caldarelli}.

We also note that there are some discussions on degree cut-off \cite{boguna}
because our analysis assumes that there exist
high-degree nodes.
However, we do not introduce such degree cut-off because we are performing
a kind of mean-field analysis.
It is to be noted that scale-free networks are a kind of random networks
and thus we can have a node with even degree $n-1$ with very small
probability if $P(n-1) > 0$ \cite{newman}.
In the mean-field analysis, such rare cases are taken into account.
However,
discussions of degree cut-off are based on average case analysis, and
there does not exist a consensus cut-off value.
Therefore, we do not introduce degree cut-off in our analysis.
The results of computer simulation support that our analysis is appropriate.
Note also that each node with degree more than 1 must be
covered by $C$ nodes (not by $C$ edges) in our ILP formulation and thus
the effect of multiple edges is eliminated in computer simulation. 
As it has been shown in the field of complex network science, the analysis and classification of
networks in terms of their degree distribution is a key feature to understand the complex behavior of complex systems. 
In particular, the scale-free topology fundamentally changes 
the system's behavior, with broad implications from spreading processes on networks 
(like for example the spread of infectious diseases) to cascading failures \cite{newman, baraliu, caldarelli}. It is therefore appropriate to examine the
controllability problem in networks governed by power-law degree distributions.

First we assume that the minimum degree $D$ is 2 in an undirected graph $G(V,E)$,
where
$V$ is a set of $n$ nodes, and $E$ is a set of edges connecting nodes in $V$. 
From the following equation:
\[
\alpha n \int_{2}^n k^{-\gamma}  dk
= {\frac {\alpha n}{\gamma-1}} \cdot
\left( {\frac 1 {2^{\gamma-1}}} - {\frac 1 {n^{\gamma-1}}} \right)
\approx {\frac {\alpha n}{\gamma-1}} \cdot {\frac 1 {2^{\gamma-1}}}
= n,
\]
we have $\alpha = 2^{\gamma-1}(\gamma-1)$.

Let $DS$ be the set of nodes with degree between $n^{\beta}$ and $n$.
Then, the number of nodes in $DS$ (denoted by $N_{DS}$) is
\[
N_{DS} \approx  \alpha n \int_{n^{\beta}}^{n} k^{-\gamma} dk 
=  - 2^{\gamma-1} n \cdot \left[ k^{1-\gamma} \right]_{n^{\beta}}^{n} 
= - 2^{\gamma-1} n \cdot \left( n^{1-\gamma} - n^{\beta(1-\gamma)} \right) 
\approx  2^{\gamma-1} n^{1+\beta(1-\gamma)}.
\]

Let $E_G$ be the number of edges in $G(V,E)$.
Then, $E_G$ is given by
\[
E_{G} \approx  {\frac {\alpha n} 2} \cdot \int_{2}^n k \cdot k^{-\gamma} dk \approx 
{\frac {\alpha n} {2 (2 - \gamma)}} \cdot n^{2-\gamma},
\]
where the factor 2 in ${\frac {\alpha n} 2}$ comes from the fact that
each edge is counted by two nodes.
The number of edges that are connected to at least one node in $DS$
(i.e., the number of edges covered by $DS$)
is lower bounded by
\[
E_{DS} =
{\frac {\alpha n}{2}} \cdot \int_{n^{\beta}}^n k \cdot k^{-\gamma} dk =
{\frac {\alpha n}{2 (2 - \gamma)}} \cdot (n^{2-\gamma}-n^{\beta(2-\gamma)}).
\]
It should be noted that $E_{DS}$ gives a lower bound and
the number of edges covered by $DS$ may be much larger 
because this estimate considers the case where
both endpoints of these edges are in $DS$.

The probability that an arbitrary edge is not covered by $DS$ is
upper bounded by
\[
{\frac {E_G - E_{DS}} {E_G}} \approx {\frac {n^{\beta(2-\gamma)}}{n^{2-\gamma}}}
= n^{(\beta-1)(2-\gamma)}.
\]
Then, the probability that a node with degree $k$ does not have
any edge connected to $DS$ is upper bounded by
\[
n^{k(\beta-1)(2-\gamma)},
\]
which is also upper bounded by $n^{2(\beta-1)(2-\gamma)}$ because
the minimum degree is assumed to be 2.
Therefore, the number of nodes (denoted by $N_{G-DS}$)
not covered by $DS$ is
\[
N_{G-DS} \leq O(n \cdot n^{2(\beta-1)(2-\gamma)}) = O(n^{1+2(\beta-1)(2-\gamma)}).
\]
Since we can have a dominating set if we merge these nodes with $DS$,
the number of nodes in an MDS is upper bounded by $N_{DS}+N_{G-DS}$.
To minimize the order of $N_{DS} + N_{G-DS}$.
we let
\[
1+\beta(1-\gamma) = 1+2(\beta-1)(2-\gamma)
\]
which results in
\[
\beta = {\frac {2 (2 - \gamma)}{3 - \gamma}}.
\]
By using this $\beta$,
an upper bound of the size of an MDS is estimated as
\[
O(n^{1-{\frac {2 (\gamma-1)(2-\gamma)}{3-\gamma}}}).
\]

We can see that this order is smaller than that of our previous result on $D$=1
\cite{nacher2}
\[
O(n^{1-(\gamma-1)(2-\gamma)}).
\]
In particular, the above takes the minimum order $O(n^{0.75})$ when
$\gamma^*=1.5$ for $D$=1 whereas the new bound for $D$=2 takes the minimum order $O(n^{0.657})$
when $\gamma^*=3-\sqrt{2}$ (see Fig. 2ab).
This difference comes from the fact that a node $v$ is regarded as
not covered by $DS$ if one specific edge connected to $v$ is not covered by $DS$
in an existing analysis \cite{nacher2}
whereas
a node $v$ is regarded as not covered by $DS$ if no edge connected to
$v$ is not covered by $DS$ in this analysis.


We can extend the above result for the case where the minimum degree is $D$,
by replacing
\[
N_{G-DS} \leq O(n \cdot n^{2(\beta-1)(2-\gamma)}) = O(n^{1+2(\beta-1)(2-\gamma)}),
\]
with
\[
N_{G-DS} \leq O(n \cdot n^{D(\beta-1)(2-\gamma)}) = O(n^{1+D(\beta-1)(2-\gamma)}).
\]
Then, we have
\begin{eqnarray}
1+\beta(1-\gamma) & = & 1+D(\beta-1)(2-\gamma),\\
\beta & = & {\frac {D(2-\gamma)}{D(2-\gamma)+(\gamma-1)}}.
\end{eqnarray}
By using this $\beta$,
an upper bound of the size of MDS is estimated as
\begin{eqnarray}
O(n^{1-{\frac {D(2-\gamma)(\gamma-1)}{D(2-\gamma)+\gamma-1}}}).
\end{eqnarray}
This order of the MDS size that scales as $n^{\delta}$ takes the minimum value when
\[
\gamma^* = {\frac {(2D-1)-\sqrt{D}}{D-1}}.
\]
It is to be noted that although $\alpha$ depends on $D$,
it does not affect the order of the MDS size. The scaling exponent $\delta$ for the order of the MDS size is shown as a function 
of the degree exponent $\gamma$ in Fig. 2d. This is our first main result and demonstrates that 
for scale-free networks with $\gamma<2$, the order of the MDS size changes 
(the exponent $\delta$ changes in functional form of $n^{\delta}$) when the minimum degree increases. 
The dependence 
of the degree exponent $\gamma^*$ that minimises the MDS size on the minimum degree $D$ is also shown in
 Fig. 2c. The 
results demonstrate that a higher minimum degree makes it easier to control scale-free networks with $\gamma<2$ (Fig. 2d). 

\begin{figure}[th]
\includegraphics[angle=-90, width=16cm]{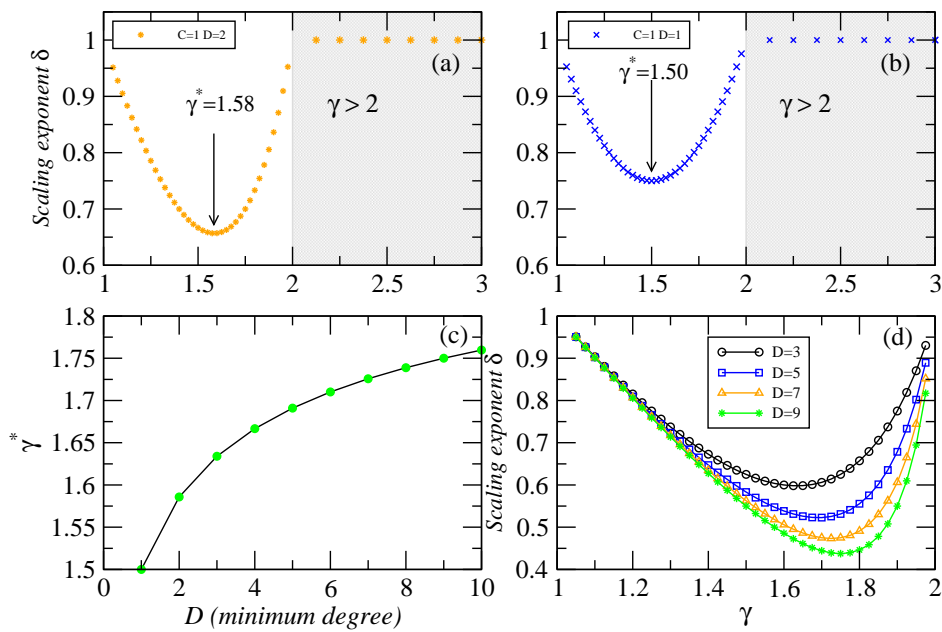}
\caption{{\bf The theoretical predictions for the MDS order.} 
Theoretical analysis illustrates that the MDS size scales according to $n^{\delta}$ (see Eq. 3) in scale-free networks of size $n$. The scaling exponent 
$\delta$ as a function of the degree exponent $\gamma$ for the
(a) $C=1$, $D=2$ and (b) $C=1$, $D=1$ configurations.
(c) The degree exponent that minimises the MDS as a function of the minimum degree $D$. (d) The dependence of the $\delta$ exponent 
on the degree exponent $\gamma$ calculated for several $D$ values.}
\label{fig:theory}
\end{figure}

\subsection {Analysis on robust domination (RMDS) with minimum degree $D$ and a generic $C$-cover.}
\subsubsection{Analysis for the case of $\gamma<2$}

Next, we show the results for the robust domination (RMDS) (Fig. 1d). For an undirected graph $G(V,E)$ and a positive integer $C$,
$RDS \subseteq V$ is called a $C$-robust dominating set if
each node $v \in V$ satisfies the following:
either $v \in RDS$ or $v$ is connected to $C$ or more nodes in $RDS$.
Here, we provide an upper bound of the size of the minimum $C$-robust dominating set. Note that an RDS 
is a special case of a generalized dominating set \cite{add1, add2}, which has been
studied in the context of the computational complexity. However, it has not been investigated from the perspective
of complex networks. We consider the case of $C$-robust domination in which 
the minimum degree of $G(V,E)$ is $D$,
where $C$ and $D$ are constants such that $D \geq C$.


As in Section II.A,
let $DS$ be the set of nodes with degree between $n^{\beta}$ and $n$.
Then, the probability that a node is not covered by $C$ or more nodes
in $DS$ is bounded by
\[
O \left( \sum_{k=D-C+1}^{D} \left(
\begin{array}{c}
D \\
k
\end{array}
\right) n^{(\beta-1)(2-\gamma)k} 
\right)
\]
where we do not include the factor of
$(1-O(n^{(\beta-1)(2-\gamma)})^{D-k}$ because
we consider an upper bound.
This number is further bounded by
\[
O( D \cdot D^{D-C+1} \cdot n^{(\beta-1)(2-\gamma)(D-C+1)} )
\]
for sufficiently large $n$.
Therefore, the number of nodes not covered by $DS$ is
\[
O(n \cdot n^{(\beta-1)(2-\gamma)E}) = O(n^{1+E(\beta-1)(2-\gamma)}),
\]
where we let $E = D-C+1$.
It is to be noted that a constant factor is ignored here because
we use $O$-notation.

As before, by balancing the size of $DS$ and the number of non-covered nodes,
we have
\begin{eqnarray}
1+\beta(1-\gamma) & = & 1 + E(\beta-1)(2-\gamma),\\
\beta & = & {\frac {E(2-\gamma)}{E(2-\gamma) + (\gamma-1)}}.
\label{eq:degDcovC}
\end{eqnarray}
By using this $\beta$, an upper bound of the size of RMDS is estimated as
\begin{eqnarray}
O(n^{1-{\frac {E(2-\gamma)(\gamma-1)}{E(2-\gamma)+\gamma-1}}}).
\end{eqnarray}

This is our second and most important finding. This result suggests that the case of an RMDS with
minimum cover $C$ and minimum degree $D$
corresponds to the case of an MDS with the minimum degree $D-C+1$.
For example, the case of an RMDS with $C=D=2$ (i.e., the case where each node (with a degree of at least 2)
must be covered twice, and the minimum degree $D$ is 2) corresponds to
the case of an MDS with $D=1$.


In all theoretical analyses in Sections II.A and II.B,
we assume that multi-edges (i.e., multiple edges between the same pair
of nodes) are allowed
because it is known that there does not exist a network strictly
following a power-law distribution with $\gamma <2$
if multi-edges are not allowed \cite{delgenio}.
However, even if multi-edges between the same pairs are replaced by
single edges after generating a power-law network with multi-edges,
the results should hold if the cover parameter $C$ is 1
because MDS is only concerned with existence of
an edge from each node not in MDS to a node in MDS.
If $C>1$,
we need to consider the possibility that some of $C$ edges are connected 
to a node $v$ are
multi-edges because such $v$ may not be dominated by $C$ nodes. 
We will show that such a factor can be ignored in many cases
if we discuss the order of the size of MDS.
Of course, the resulting network does not strictly follow a power-law
distribution
if multi-edges are replaced by single edges.
However, because any network with $\gamma < 2$ cannot strictly follow a power-law
distribution,
our assumption seems reasonable.

Here we note that Eqs.~(4)(5) are identical to Eqs.~(1)(2)
if we replace $E$ by $D$.
It suggests that the case of RMDS with
the minimum cover $C$ and the minimum degree $D$
corresponds to the case of MDS with the minimum degree $D-C+1$.
For example, the case of RMDS with $C=D=2$ corresponds to
the case of MDS with $D=1$.

In the above, we implicitly assumed that all $C$ edges are connected
to different nodes in $DS$.
However, we need to consider the possibility that 
some of $C$ edges are connected to the same node in $DS$ because
we allow multi-edges in theoretical analyses.
Suppose that $C$ (or more) edges from $v$ are connected to $DS$.
Since the number of edges connected to a node of degree $k$ in $DS$ is 
$O(k)$,
there exist $O(nk^{-\gamma})$ nodes of degree $k$ in $DS$,
and there exist $O(n^{3-\gamma})$ edges connected to $DS$,
the probability that $C$ edges contain at least one common endpoint
in $DS$ is
\[
O\left( C^2 \cdot {\frac {\int_{k=1}^n k^2 (nk^{-\gamma}) dk}{(n^{3-\gamma})^2}} \right)
=
O(C^2 \cdot {\frac 1 {n^{2-\gamma}}}).
\]
Since there exist $O(n)$ nodes covered by $DS$,
the number of nodes not covered by $C$ different nodes would be
\[
O(C^2 \cdot n^{\gamma-1}).
\]
If the exponent $\gamma-1$ is smaller than that in Eq.~(6),
this factor does not affect the order of Eq.~(6). For $\gamma < 1.5$, it is true for $E \leq 10$.
However, if $\gamma \geq 1.7$,
it is true only for $E=1$ (i.e., $C=D$).
Therefore, we need to be careful if we consider
the case of $D>C$ and $\gamma \geq 1.7$.

\subsubsection{Analysis for the case of $\gamma > 2$}
\label{sec:gsmall}

\paragraph{Analysis of lower bound}

First we consider a lower bound.
Let $D$ be the minimum degree.
From $\alpha n \int_{D}^{n} k^{-\gamma} dk ~=~ n$,
we have $\alpha = ({\gamma-1}) \cdot D^{\gamma-1}$.

For $S \subseteq V$,
$\Gamma(S)$ denotes the set of edges between $S$ and $V-S$.
(i.e., $\Gamma(S) = \{ \{u,v\}~|~u \in S \mbox{ and } v \in V-S \}$).
Here we assume without loss of generality that $|S| < n/2$ because
we are only interested in cases where $|S|$ is small compared with $n$.
The following property is trivial
\begin{eqnarray}
\mbox{if } |\Gamma(S)| < n/2, ~~S \mbox{ can not dominate $V$}.
\label{eq:nodom}
\end{eqnarray}

Let $S$ be the set of nodes whose degree is greater than or equal to $K$.
We estimate the size of $\Gamma(S)$ as follows.
\begin{eqnarray*}
|\Gamma(S)| & < & \alpha n \int_{K}^{n} k \cdot k^{-\gamma} dk
~\approx~  n (\gamma-1)D^{\gamma-1} \int_{K}^{n} k^{-\gamma+1} dk \nonumber \\
& = & n \cdot D^{\gamma-1} \cdot \left({\frac {\gamma-1}{\gamma-2}}\right) \cdot \left( {\frac 1 {K^{\gamma-2}}} - {\frac 1 {n^{\gamma-2}}} \right)
~<~ n \cdot D^{\gamma-1} \cdot \left({\frac {\gamma-1}{\gamma-2}}\right) \cdot {\frac 1 {K^{\gamma-2}}}.
\label{eq:gs1}
\end{eqnarray*}
If $S$ is a dominating set, the last term should be no less than $n/2$.
Therefore, the following inequality should be satisfied:
\begin{eqnarray}
n \cdot D^{\gamma-1} \cdot \left({\frac {\gamma-1}{\gamma-2}}\right) \cdot {\frac 1 {K^{\gamma-2}}} ~>~ n/2 .
\label{eq:domcond}
\end{eqnarray}
By solving this inequality, we have
\begin{eqnarray*}
K ~<~ \left[ D^{\gamma-1} \cdot \left({\frac {\gamma-1}{\gamma-2}} \right)
\cdot \left( {\frac {n}{n/2}} \right)
\right]^{1/(\gamma-2)} ~=~
\left[ 2 D^{\gamma-1} \cdot \left({\frac {\gamma-1}{\gamma-2}} \right)
\right]^{1/(\gamma-2)}.
\end{eqnarray*}
Then, the size of $S$ is estimated as
\begin{eqnarray*}
|S| & \approx &  \alpha n \int_{K}^{n} k^{-\gamma} dk
~\approx~ n \left( {\frac 1 {K^{\gamma-1}}} - {\frac 1 {{n}^{\gamma-1}}} \right)
~\approx~ n \cdot {\frac 1 {K^{\gamma-1}}} \nonumber \\
& > &  \left[ 2 D^{\gamma-1} \cdot \left({\frac {\gamma-1}{\gamma-2}} \right) \right]^{-{\frac {\gamma-1} {\gamma-2}}} \cdot n.
\label{eq:lb}
\end{eqnarray*}

We extend the above analysis to $C$-domination (i.e., each node must be
covered by $C$ or more edges).
In this case, Ineq. (\ref{eq:nodom}) should be replaced by
\begin{eqnarray*}
\mbox{if } |\Gamma(S)| < nC/2, ~~S \mbox{ can not $C$-dominate $V$}.
\label{eq:nodomC}
\end{eqnarray*}
Then, Ineq.~(\ref{eq:domcond}) is also replaced by
\begin{eqnarray*}
n \cdot D^{\gamma-1} \cdot \left({\frac {\gamma-1}{\gamma-2}}\right) \cdot {\frac 1 {K^{\gamma-2}}} ~>~ (nC)/2 .
\label{eq:domcondC}
\end{eqnarray*}
Finally, we have
\begin{eqnarray}
|S| ~>~  \left[ \left( D^{\gamma-1} \cdot {\frac 2 C} \right)
\cdot \left({\frac {\gamma-1}{\gamma-2}} \right) \right]^{-{\frac {\gamma-1} {\gamma-2}}} \cdot n ~=~
\left( {\frac C {2 D^{\gamma-1}}} \right) ^{\frac {\gamma-1}{\gamma-2}} 
\cdot \left[ \left({\frac {\gamma-1}{\gamma-2}} \right) \right]^{-{\frac {\gamma-1} {\gamma-2}}} \cdot n .
\label{eq:lbC}
\end{eqnarray}
For example, consider the case of $C=2$ and $\gamma = 3$ for fixed $D$.
In this case, 2-domination requires $2^2=4$ times larger MDS.

%
%
%
\paragraph{Analysis of upper bound}

Next we consider an upper bound.
As in the above,
we have $\alpha = ({\gamma-1}) \cdot D^{\gamma-1}$.
Let $DS$ be the set of nodes with degree between $B$ and $n$.
Then, the size of $DS$, $N_{DS}$, is estimated as
\[
N_{DS} \approx  \alpha n \int_{B}^{n} k^{-\gamma} dk 
=  - n D^{\gamma-1}\cdot \left[ k^{1-\gamma} \right]_{B}^{n} 
= - n D^{\gamma-1} \cdot \left( n^{1-\gamma} - B^{1-\gamma} \right) 
\approx  n \cdot {\frac {D^{\gamma-1}} {B^{\gamma-1}}}.
\]

As in Section~\ref{sec:gsmall},
let $E_G$ and $E_{DS}$ be the number of edges in $G(V,E)$ and
the number of edges connected to $DS$, respectively.
In addition, let $D$ denote the minimum degree.
Then, we have
\begin{eqnarray*}
E_{G} & \approx  & \alpha n \int_{D}^n k \cdot k^{-\gamma} dk \approx 
{\frac {\gamma-1}{\gamma-2}} \cdot
n \cdot {\frac {D^{\gamma-1}}{D^{\gamma-2}}},\\
E_{DS} & \approx & \alpha n \int_{B}^n k \cdot k^{-\gamma} dk \approx
{\frac {\gamma-1}{\gamma-2}} \cdot
n \cdot {\frac {D^{\gamma-1}} {B^{\gamma-2}}}.
\end{eqnarray*}
The probability that an arbitrary edge is covered by $DS$ is
\begin{eqnarray*}
{\frac {E_{DS}}{E_G}} \approx \left( {\frac D B} \right)^{\gamma-2}.
\end{eqnarray*}
Thus, a lower bound of the probability that an arbitrary node
is not covered by $C$ or more edges is estimated as
\begin{eqnarray*}
1 - \sum_{k=C}^{D} 
\left(
\begin{array}{l}
D \\
k
\end{array}
\right)
\left[ \left( {\frac D B} \right)^{\gamma-2} \right]^k
\left[ 1- \left( {\frac D B} \right)^{\gamma-2} \right]^{D-k}
~=~
\sum_{k=0}^{C-1} 
\left(
\begin{array}{l}
D \\
k
\end{array}
\right)
\left[ \left( {\frac D B} \right)^{\gamma-2} \right]^k
\left[ 1- \left( {\frac D B} \right)^{\gamma-2} \right]^{D-k} .
\end{eqnarray*}
Since it is very difficult to consider a general pair $(C,D)$,
we consider the case of $C=D$.
Then, this probability is simplified into
\begin{eqnarray*}
1 - \left( {\frac D B} \right)^{C(\gamma-2)}
\end{eqnarray*}
Therefore,
an upper bound $f(B)$ of the size of MDS is estimated as
\begin{eqnarray*}
f(B) & = &
n \cdot \left[ \left({\frac D B}\right)^{\gamma-1} + 1 -
\left( {\frac D B} \right)^{C(\gamma-2)} \right] .
\end{eqnarray*}
It is to be noted that this number does not give a meaningful bound
for many $(C,\gamma)$. 
For example, if $C=1$ and $\gamma=3$, $f(B)=n$ holds.

By solving $f'(B)=0$, we see that $f(B)$ takes the minimum value
\begin{eqnarray}
n \cdot \left[ 1 +
\left({\frac {C(\gamma-2)}{\gamma-1}}\right)^{\frac {\gamma-1}{C(2 - \gamma)+(\gamma-1)}} -
\left({\frac {C(\gamma-2)}{\gamma-1}}\right)^{\frac {C(\gamma-2)}{C(2 - \gamma)+(\gamma-1)}}  \right]
\end{eqnarray}
at
$
B  = D \cdot
\left( {\frac {\gamma-1}{C(\gamma-2)}} \right)^{1/(C(2-\gamma)+(\gamma-1))}
$.
It is interesting to note that
this minimum value does not depend on the minimum degree $D$.
If $C=D=1$ (i.e., original MDS),
this minimum value is simplified into
\begin{eqnarray*}
n \cdot \left[ 1
+ \left({\frac {\gamma-2}{\gamma-1}}\right)^{\gamma-1}
- \left({\frac {\gamma-2}{\gamma-1}}\right)^{\gamma-2}
\right].
\end{eqnarray*}

It is also interesting to consider the case of $C=1$ and $D=2$.
In this case, $f(B)$ is given by
\begin{eqnarray*}
f(B) & = &
n \cdot \left\{ \left({\frac 2 B}\right)^{\gamma-1} + 
\left[ 1 - \left( {\frac 2 B} \right)^{(\gamma-2)} \right]^2 \right\} .
\end{eqnarray*}
Although it is difficult to analytically derive its minimum,
we can estimate it by numerical computation.
Fig. 3 compares upper bounds for $(C,D)=(1,1)$
and $(C,D)=(1,2)$ and lower bounds for $(C,D)=(1,1)$. This figure shows that the upper
bound becomes smaller as $D$ increases in the case of $C=1$.

\begin{figure}[th]
\begin{center}
\includegraphics[width=12cm]{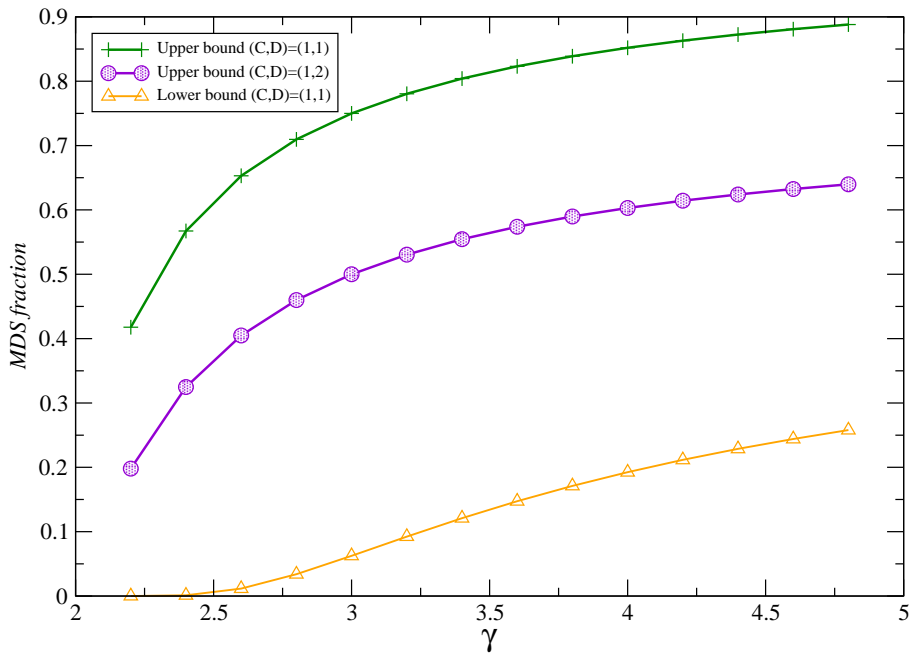}
\caption{{\bf Analytical results for several configurations.} Comparison of upper bounds for $(C,D)=(1,1),(1,2)$ and
lower bounds for $(C,D)=(1,1)$.}
\label{fig:ublb}
\end{center}
\end{figure}

Although, as shown in Fig. 4, the gap between the derived lower and upper bounds is large (especially for larger $C$ and $D$), these are not trivial. For example,
suppose that there exist two nodes with degree no less than $n-1$. Then, the size of RMDS for $C=2$ is 2, which is $O(1)$
(much less than $O(n)$). The $\Omega(n)$ lower bound suggests that such a case seldom occurs in random scale-free networks.

\section{Computation of robust domination (RMDS)}

\subsection{ILP-formulation for MDS in unipartite networks.} 
Let $G(V,E)$ be an undirected graph,
where $V$ and $E$ are sets of nodes and edges, respectively.
We begin with ILP (Integer Linear Programming)-formulation for computation of an MDS \cite{nacher1}.
From $G(V,E)$, we construct the following ILP instance:
\begin{eqnarray*}
\mbox{minimize} & ~ & \sum_{v \in V} x_v, \\
\mbox{subject to} & ~ & x_u + \sum_{\{u,v\}\in E} x_v \geq 1
~~~\mbox{for all $u \in V$},\\
& ~ & x_v \in \{0,1\} ~~~\mbox{for all $v \in V$}.
\end{eqnarray*}
Then, the set $\{v | x_v = 1\}$ clearly gives an MDS. It is known that,
in contrast to the bipartite matching \cite{baraliu}, the MDS problem
is NP-hard. Therefore, it is reasonable to use ILP. 

\subsection{ILP-formulation for robust domination (RMDS)}

Suppose that each node must be covered twice except degree 1 and 0 nodes.
Then, we can formulate this \emph{robust dominating set} problem for $C=2$ (i.e.,
each node (with degree greater than 2) is either in MDS
or is covered by at least two nodes in MDS, where each node
with degree 1 is either
in MDS or is covered by at least one node in MDS) as follows.
Because it is impossible to cover each degree 1 node by two edges, we have introduced
this exceptional handling of degree 1 nodes. However, if the minimum degree is 2 or more, we need not consider
this exceptional case.

\begin{eqnarray*}
\mbox{minimize} & ~ & \sum_{v \in V} x_v,\\
\mbox{subject to} &  & 2 \cdot x_u ~+~ \sum_{\{v,u\} \in E} x_v ~\geq~ 2 ~~~~\mbox{for all $u \in V$ such that $deg(u)>1$},\\
&  & x_u ~+~ \sum_{\{v,u\} \in E} x_v  ~\geq~ 1 ~~~~\mbox{for all $u \in V$ such that $deg(u)=1$},\\
& & x_v \in\{0,1\} ~~~~\mbox{for all $v \in V$.}
\end{eqnarray*}

However, {\bf glpsol} (GNU Linear Programming Solver) executable could not solve this problem in reasonable CPU time. 
So, we strengthen the condition so that
each node with degree greater than 2 is covered by at least two nodes in MDS
even if the node belongs to MDS.
Then, the resulting IP becomes as follows.

\begin{eqnarray*}
\mbox{minimize} & ~ & \sum_{v \in V} x_v,\\
\mbox{subject to} &  & x_u ~+~ \sum_{\{v,u\} \in E} x_v ~\geq~ 2 ~~~~\mbox{for all $u \in V$ such that $deg(u)>1$},\\
&  & x_u ~+~ \sum_{\{v,u\} \in E} x_v ~\geq~ 1 ~~~~\mbox{for all $u \in V$ such that $deg(u)=1$},\\
& & x_v \in\{0,1\} ~~~~\mbox{for all $v \in V$.}
\end{eqnarray*}



It is to be noted that the solution obtained by the above ILP also satisfies the conditions of the original formulation. Therefore,
the solution obtained by this ILP also gives a robust dominating set although it is not necessarily minimum.
We can also consider a variant of MDS in which
weight $w(u,v)$ is assigned for each edge and
each node $u$ must be covered by
edges with total weight $W_u$.
Then, this variant can be formulated as 

\begin{eqnarray*}
\mbox{minimize} & ~ & \sum_{v \in V} x_v,\\
\mbox{subject to} &  & w(u,u) \cdot x_u ~+~ \sum_{\{v,u\} \in E} (w(u,v) \cdot x_v) ~\geq~ W_u ~~~\mbox{for all $u \in V$ such that $deg(u)>0$},\\
& & x_v \in\{0,1\} ~~~~\mbox{for all $v \in V$.}
\end{eqnarray*}

\subsection{Implementation of the ILP problems}
For the MDS ($C$=1) and RMDS ($C$=2) configurations computed in real-world and simulated networks, the optimal solution was calculated
using 'glpsol' solver (http://www.gnu.org/software/glpk). The GNU Linear Programming Kit (GLPK) supplies a software package intended 
for solving large-scale linear programming (LP), mixed integer programming (MIP), 
and other related problems. In our problem, after translating the mathematical problem into an ILP problem,
the input model is solved using GNU Linear Programming Solver (glpsol) executable.

For the probabilistic MDS (PMDS), to be shown later, the optimal solution for the ILP-formulation
was calculated using the IBM ILOG CPLEX Optimizer Studio ver.12.02. As the GLPK, it is a software package that allows
to solve large-scale mathematical optimization problems. The computation of the PMDS is more intensive than that of MDS and RMDS,
therefore we used CPLEX because it performed faster than GLPK to find the optimal solution.  

\subsection{Generation of unipartite scale-free networks}

We employ the following algorithm to construct unipartite scale-free networks of size $n$,
in which the degree distribution of $V$ (a set of nodes)
follows $P(k) \propto k^{-\gamma}$
under the constraint that the minimum and maximum degrees are $D$ and $n$,
respectively.

For given $n,\gamma,D$ we generate a random unipartite network
in the following way.

\begin{itemize}
\item [(1)] For each node $v \in V$, generate half edges $e_i = (v, u_i)$
($u_i$ is a virtual node) according to the degree distribution
$\alpha_1 k^{-\gamma}$ under the constraint of
the minimum degree $D$ and the maximum degree $n$,
where $\alpha$ is selected so that the number of nodes in $V$ is almost $n$.
\item [(2)] Repeat the following
until there are almost no remaining half edges: randomly select non-connected
$e_i = (v,u_i)$ and $e_j = (v',u_j)$ such that $v \neq v'$ and then connect $v$ and $v'$.
\end{itemize}


The probabilistic MDS, to be introduced later, was computed using generated samples of synthetic scale-free networks 
with a variety of scaling exponent $\gamma$ and average degree $<k>$ values using the Havel-Hakimi algorithm 
with random (Monte-Carlo) edge swaps (HMC) \cite{HHM}. 

\subsection{Computer simulations for RMDS}

To confirm the theoretical predictions shown above, we constructed artificial scale-free networks with a variety of degree 
exponents $\gamma$ and minimum degree $D$=1 and $D$=2. An ensemble of scale-free networks was constructed
for each network size up to 10,000 nodes,
and the mean value together with standard error of the mean (s.e.m.) 
for MDS size with $C$=1 and $C$=2 were computed. For $\gamma<2$, the theoretical results 
predict the same order of MDS size (the same exponent $\delta$ in the scaling function $n^{\delta}$) 
for configurations ($D$=2, $C$=2) and ($D$=1, $C$=1) (see Eq. 6). In contrast, the results predict 
a different scaling for the configuration ($D$=2, $C$=1), as shown by Eq. 3. Fig. ~\ref{fig:combined_sim} presents
the simulation results 
for $\gamma<2$, which agree with the analytical predictions. 

For $\gamma>2$, the analytical computations predict the same scaling functional 
form $n^{\delta}$ with ${\delta=1}$ for all the three configurations. 
The simulation results agree with this prediction with high accuracy (see Figs. 6 and 7). 

\subsection{Robust control of real-world networks}

We used the concepts and mathematical tools presented above to investigate the robust control of 
several real networks. The experimental data analysis includes undirected, directed and bipartite networks 
from biological and socio-technical systems (see Tables 1, 2 and 3). We first present the results 
for undirected networks and show that the MDS density for $C$=1 increases with increasing $\gamma$. The computation of the robust 
MDS density ($C$=2) exhibits a similar dependency, as predicted by Eq. 9 (Fig. 8ab). Interestingly,  
the MDS ratio for $C$=2 and $C$=1
differs, on average, by a factor of 2 or less (Fig. 9d), in agreement 
with the theoretical predictions shown in Fig. 4c. 
When overlapping the real 
data and the predictions from Eqs. 9 and 10 for the lower and upper bounds, respectively, for networks with $\gamma>2$, we see that 
the real data are always within the theoretical boundaries (Fig. 10).

The MDS size for both  $C$=1 and $C$=2 scales linearly with $n$ (see Fig. ~\ref{fig:2x3}a), which is in agreement with 
the theoretical predictions shown in Eqs. 9 and 10 for 
$\gamma>2$ and the computer simulations (see Figs. 6-7). Note that Fig. 8 
displays the MDS fraction,
and Fig. ~\ref{fig:2x3}a represents the MDS size.

\begin{figure}[th]
\includegraphics[angle=-90, width=16cm]{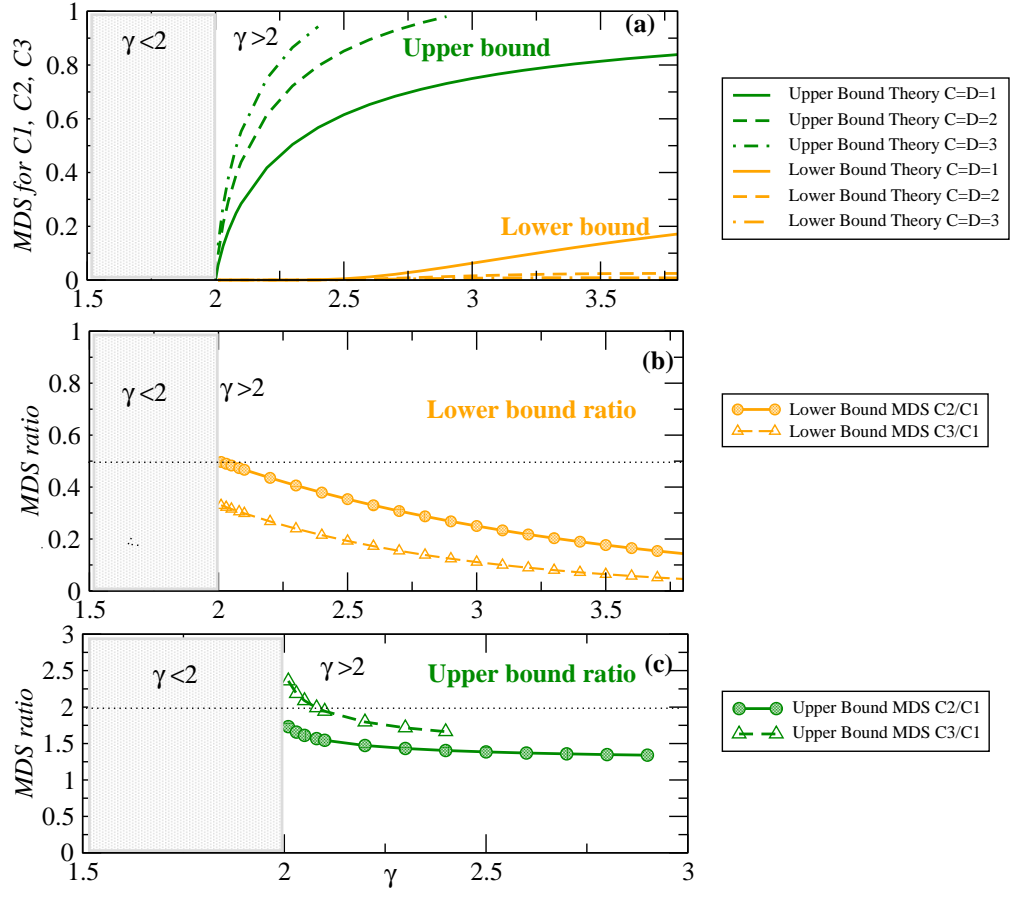}
\caption{{\bf The theoretical predictions for MDS($C$=1) and robust domination RMDS($C=2,3$) with $D=C$.} (a) Lower and 
upper bound predictions for the fraction of nodes required to control the entire network 
for covers $C=1,2,3$. (b) Ratio of lower bound MDS sizes for $C=2$ and $C=1$ (circles) and $C=3$ and $C=1$ (triangles). 
(c) Ratio of upper bound MDS sizes for $C=2$ and $C=1$ (circles) and $C=3$ and $C=1$ (triangles).}
\label{fig:theory_bounds}
\end{figure}

To investigate the influence of the frequency of nodes with degree 1 and 2 ($P(1)=n(1)/n$ and $P(2)=n(2)/n$) on robust control, we 
computed MDS1 and MDS2 versus $P(1)$ and $P(2)$. We then calculated the ratio of $MDS2/MDS1$ versus $P(1)$ and $P(2)$. The 
results indicate that a small $P(1)$ and large $P(2)$ tend to be associated with a small MDS density 
(see Fig. 11). The ratio of MDS2/MDS1 is less than 2 in most cases. 

The analysis for directed networks included an Internet peer-to-peer (P2P) network, the transcriptional regulatory network for {\it E. coli} 
from the Regulon database, a set of food webs from different ecosystems, U.S. political blogs and the chemical synapse 
network for {\it C. elegans}.
The results demonstrate that MDS1 and MDS2 densities increase with increasing
 $\gamma_{out}$ (Figs. 12a-b) and $\gamma_{in}$ (Fig. 13a-b), which is in agreement with the dependence found 
for undirected networks. In addition, the MDS sizes for $C$=1 and $C$=2 scale linearly with $n$ (Fig. 9bc). 
Moreover, as in the undirected case, 
the MDS ratio between $C$=1 and $C$=2 is almost always less than two, with only one exception (see Fig. 9d-f). Moreover,
less than 50$\%$ of nodes are needed to control the network in both the typical ($C$=1) and robust ($C$=2) control 
configurations (see Figs. 12c-d and 13c-d). 


\section{Analysis on robust domination (RMDS) in bipartite networks}

\subsection{Computation of MDS in bipartite networks}
We define a bipartite graph as $G(V_{\top},V_{\bot};E)$, where
$V_{\top}$ is a set of top nodes, $V_{\bot}$ is a set of bottom nodes,
and $E$ is a set of edges ($E \subseteq V_{\top} \times V_{\bot}$). In our analysis, 
the directions of the edges are considered 
from $V_{\top}$ to $V_{\bot}$. Therefore, the set of driver nodes 
will be a subset of $V_{\top}$, where nodes in $V_{\top}$ need not be covered.

The computation of an MDS of a bipartite network
is equivalent to the computation of a minimum set cover. 
Although it is an NP-hard problem, we have verified
that the optimal solution is obtained in networks with power-law distributions 
of up to approximately 110,000 nodes within a few seconds.
The computation was formalised as the following Integer Linear Programming (ILP) problem 
\begin{equation}
	\begin{aligned}
\text{minimize} &  \sum_{v \in V_{\top}} x_v,\\
\text{subject to} & \sum_{\{v,u\} \in E} x_v \geq 1 \hspace{0.5cm} \text{for all} \hspace{0.5cm} u \in V_{\bot}, \\
& x_v \in {0,1} \hspace{0.5cm} \text{for all} \hspace{0.5cm} v \in V_{\top}. 
	\end{aligned}
\end{equation}


\subsection{Computation of RMDS in bipartite networks}

The above mentioned ILP can be extended for computation of
an RMDS in bipartite networks.
It is formalized as
\begin{eqnarray*}
\mbox{minimize} & ~ & \sum_{v \in V_{\top}} x_v,\\
\mbox{subject to} &  &  \sum_{\{v,u\} \in E} x_v \geq 1
~~~\mbox{for all $u \in V_{\bot}$ such that $deg(u)=1$},\\
&  &  \sum_{\{v,u\} \in E} x_v \geq 2
~~~\mbox{for all $u \in V_{\bot}$ such that $deg(u)>1$}.
\end{eqnarray*}
where $deg(u)$ indicates the degree of node $u$. It should be noted that for any node $u \in V_{\bot}$ with degree 1,
it is not possible to cover $u$ twice and thus we must relax the condition
for these nodes.

\subsection{Generation of bipartite scale-free networks}
We employ the following algorithm to construct bipartite scale-free networks, in which
 the degree distributions of $V_{\top}$ and $V_{\bot}$
follow $P_{\top}(k) \propto k^{-\gamma_1}$ and 
$P_{\bot}(k) \propto k^{-\gamma_2}$, respectively.
Here, we consider $n_1=|V_{\top}|$ and $n_2=|V_{\bot}|$. The maximum degree for the nodes
in $V_{\top}$ corresponds to $n_1$.

Then, for given $n_1$, $\gamma_1$, $\gamma_2$ we generate a random bipartite network in the following way.
\begin{itemize}
\item [(1)] For each node $v \in V_{\top}$,
generate half edges $e_i = (v,u_i)$ ($u_i$ is a virtual node)
according to the degree distribution $\alpha_1 k^{-\gamma_1}$  where $\alpha_1$ is selected so that the 
number of nodes in $V_{\top}$ is almost $n_1$.
\item [(2)] For each node $w \in V_{\bot}$,
generate half edges $e'_j = (u'_j,w)$ ($u'_j$ is a virtual node)
according to the degree distribution $\alpha_2 k^{-\gamma_2}$ where $\alpha_2$ is selected so that  
the number of $e'_j$s is equal to the number of $e_j$s.
\item [(3)] Randomly connect $e_i$s and $e'_j$s in a one-to-one manner.
\end{itemize}
It is to be noted that $n_2$ (the number of nodes of $V_{\bot}$) is 
determined automatically in step 2 to satisfy the condition on edge
numbers. 

\subsection{Data analysis of real-world bipartite networks}

We collected a set of 10 real-world bipartite networks corresponding to socio-technical (Fig. 14a) and biological systems (Fig. 14b). 
We then formalised and computed the MDS for the $C$=1 and $C$=2 configurations. Although the MDS with $C$=2 
is always larger than the MDS with $C$=1, the difference is proportionally very small in most cases. Figs. 14ab also illustrates 
that biological systems tend to require a larger MDS size than socio-technical systems. The computer 
simulation of ensembles of bipartite scale-free networks with a variety of degree exponents also demonstrates that 
$C$=1, $D$=1 and $C$=2, $D$=2 can control the network with a similar fraction of nodes. Therefore, 
robustly controlling a bipartite network requires a similar fraction of nodes as the typical, non-robust system (see Figs. 14cd).

\section{Theoretical Analysis for the probabilistic domination (PMDS)}

In some real networks, each link has a probability of failing, which leads to the probabilistic concept of robust control (PMDS). For 
example, experimental analyses on neural networks 
have confirmed the unreliability of central synaptic transmission in rat brains \cite{failure}. The mean transmission 
failure probability was found to be $p$=0.71, with a range of 0.3 to 0.95 ($w$=0.24). In this work, we used the well-studied {\it C. elegans}
neural network to investigate probabilistic robust control.
To investigate this type of systems from a theoretical perspective using 
the robust MDS approach, we assume that each edge $(v,u)$ has the probability 
of failure $P_{v,u}$ (see Fig. ~\ref{fig:schema}e). We require that each node is covered by multiple nodes 
in an MDS so that the probability that at least one edge is active is at least $\theta$. Let $S$ be a DS
Then, $S$ must satisfy

\begin{equation}
(\forall u) (1-\prod_{v \in S} P_{v,u}) ~\geq~ \theta.
\end{equation}

As we will show later, this problem can be also formalized and solved using Integer Linear Programming (ILP).


\subsection{The case of $D$=1}
First, we consider the case of $D=1$ (i.e., the minimum degree is 1)
and $w=0$.
Let $DS$ be an MDS for $G(V,E)$ for the non-probabilistic version.
Let $U$ be a set of degree 1 nodes each of which does not belong to $DS$
but is dominated by a node in $DS$.
Let $\{u,v\}$ be the only edge connecting to $v \in U$.
We can observe:
\begin{quote}
if $1 - P_{u,v} < \theta$, $v$ must be covered by itself.
\end{quote}
Therefore,
all nodes in $U$ should be added to $DS$ (in a probabilistic version)
when $\theta > 1-P_{u,v}$.
Therefore, it is expected that the MDS size increases
approximately from
$|DS|$ to $|DS|+|U|$ at around $\theta=1-p$. 
For example, consider the case of $p=0.71$.
Then, there should be great increase of the MDS size at $\theta=1-0.71=0.29$.
It shows good agreement with the simulation result (see Figs. 16-17).
\subsection{The case of $D$=2}
We can extend the above analysis to the case of $D=2$
(i.e., the minimum degree is 2) and $w=0$.
Let $DS$ be an MDS for $G(V,E)$ for the non-probabilistic version.
In this case, we consider two types of nodes of degree 2:
\begin{itemize}
\item [(a)] $v$ has one edge connecting to a node in $DS$,
\item [(b)] $v$ has two edges connecting to nodes in $DS$,
\end{itemize}
where each node does not belong to $DS$ but is dominated by a node in $DS$.
Let $U_1$ and $U_2$ be the sets of type (a) and type (b) nodes, respectively.
Then, nodes in $U_1$ should be added to $DS$ if $1-p < \theta$.
On the other hand, nodes in $U_2$ should be added to $DS$ if $1-p^2 < \theta$.
Therefore, it is expected that the MDS size increases approximately
from $|DS|$ to $|DS|+|U_1|$ at around $\theta=1-p$ and
from $|DS|+|U_1|$ to $|DS|+|U_1|+|U_2|$ at around $\theta=1-p^2$.
In the case of $p=0.71$, these two threshold values are
$0.29$ and $0.49$, in good agreement with the simulation result.
\subsection{The case of $D$=1 and $w > 0$}
Next, we consider the case of $D=1$ and $w > 0$.
As in the above,
let $DS$ be an MDS for $G(V,E)$ for the non-probabilistic version, and
let $U$ be a set of degree 1 nodes each of which does not belong to $DS$
but is dominated by a node in $DS$.
Let $e=\{u,v\}$ be the only edge connecting to $v \in U$.
Let $p+\Delta_v$ be the failure probability of this edge $e$,
where $-w \leq \Delta \leq w$.
We can observe:
\begin{quote}
if $1 - (p +\Delta_v) < \theta$, $v$ must be covered by itself.
\end{quote}
We define $U_{\Delta}$ by
\begin{eqnarray*}
U_{\Delta} = \{ v |~ v \in U,~p+\Delta_v > p+\Delta \}.
\end{eqnarray*}
Therefore,
all nodes in $U_{\Delta}$ should be added to a DS (in a probabilistic version)
where $\Delta = 1-p-\theta$.
Here, the size of $U_{\Delta}$ is estimated as
\begin{eqnarray*}
|U_{\Delta}|  \approx  \left( {\frac {\Delta + w}{2w}} \right) \cdot |U|,
\end{eqnarray*}
where $-w \leq \Delta \leq w$.
By replacing $\Delta$ with $1-p-\theta$, we have
\begin{eqnarray*}
|U_{\Delta}|  \approx  \left( {\frac {1 - p - \theta + w}{2w}} \right) \cdot |U|.
\end{eqnarray*}
Therefore, it is expected that the MDS size is approximately given by
$|DS| + \left( {\frac {1 - p - \theta + w}{2w}} \right) \cdot |U|$.
It should be noted that
${\frac {1 - p - \theta + w}{2w}}$ becomes 0 and 1 at $\theta=1-p-w$ 
and $\theta=1-p+w$, respectively.
In the case of $p=0.71$ and $w=0.29$, these two threshold values are
$0.05$ and $0.53$, in good agreement with the simulation results (see Figs. 16-17). 
This discussion can be generalized for the cases in which $\Delta_v$ does not
follow the uniform distribution.
\subsection{The case of $D$=2 and $w>0$}
Finally, we consider the case of $D=2$ and $w > 0$.
Let $v$ be a degree-2 node and 
$u_1$ and $u_2$ be the neighboring nodes to $v$.
Then, we estimate the fraction of degree-2 nodes that does not satisfy
\begin{eqnarray*}
1 - P_{u_1,v} P_{u_2,v} \geq \theta,
\end{eqnarray*}
where such a node $v$ should be added to a DS.
For that purpose, it is enough to calculate the area
shown in Fig. 15.

For $\theta$ satisfying $(p+w)(p-w) < 1 - \theta < (p+w)*(p+w)$,
we consider the region (A) whose area is given by
\begin{eqnarray*}
P_A & = & (p+w) \cdot \left[ (p+w)- \left( {\frac {1-\theta}{p+w}} \right) \right] 
- \int_{{\frac {1-\theta}{p+w}}}^{p+w} {\frac {1-\theta}{x}} dx\\
& = &
(p+w) \cdot \left[ (p+w)- \left( {\frac {1-\theta}{p+w}} \right) \right] 
- (1-\theta) \cdot \left[ \ln(p+w) - \ln\left( {\frac {1-\theta}{p+w}} \right) \right],
\end{eqnarray*}
Therefore, in this case, the fraction is given by $P_A / ((2*w)^2)$ because
$(P_{u_1,v},P_{u_2,v})$ is uniformly distributed in the region of
$[p-w,p+w] \times [p-w,p+w]$.

For $\theta$ satisfying $(p-w)(p-w) < 1 - \theta < (p+w)*(p-w)$,
we consider the region (B) whose area is given by
\begin{eqnarray*}
P_B & = & (2w)^2 - \int_{p-w}^{{\frac {1-\theta}{p-w}}} {\frac {1-\theta}{x}} dx
+ (p-w) \cdot \left[ \left( {\frac {1-\theta}{p-w}} \right) - (p-w) \right]\\
& = & (2w)^2 - (1-\theta) \cdot \left[ \ln\left( {\frac {1-\theta}{p-w}} \right) - \ln(p-w) \right]
+ (p-w) \cdot \left[ \left( {\frac {1-\theta}{p-w}} \right) - (p-w) \right].
\end{eqnarray*}
Again, the fraction is given by $P_B / ((2*w)^2)$. The simulation results with 1,000 nodes
for $D$=1, $D$=2 with $w>0$ configurations are shown in Figs. 16-17.
We also compared the theoretical results for the case $D=2$ with those from the simulations performed
on scale-free networks with $D=2$ and $\gamma=3$. The plots show a similar overall tendendy although the inflection point
is more noticiable in the theoretical curve (see Fig. 18). It is worth noticing that 
the theoretical values $P_A$ and $P_B$ are scaled
so that these take almost the same values as the simulated ones at the
beginning and ending points (i.e., so that the values take between
0.3 and 0.85 instead of between 0.0 and 1.0)
since it is assumed in theoretical analysis that all nodes are of
degree 2 and the effects of the other nodes are ignored
(note also that degree 2 nodes occupy a major portion of nodes in the
case of $D=2$).
This comparison result suggests that theoretical analysis
captures some tendency even if nodes with degree more than 2 are
ignored.


\subsection{ILP-formulation for probabilistic robust domination (PMDS)}
\label{sec:probdom}

We assume that each edge $(v,u)$ has the probability of failure $P_{v,u}$.
We want each node be covered by multiple nodes in MDS so that
the probability that at least one edge is active is at least $\theta$.
Let $S$ be a dominating set.
Then, we require $S$ to satisfy
\[
(\forall u)
(1-\prod_{v \in S} P_{v,u}) ~\geq~ \theta.
\]
Then, we have
\begin{eqnarray*}
1-\prod_{v \in S} P_{v,u}  & \geq & \theta, \\
\prod_{v \in S} P_{v,u}  & \leq & 1-\theta, \\
\sum_{v \in S} \ln(P_{v,u})  & \leq & \ln(1-\theta), \\
\sum_{v \in S} -\ln(P_{v,u})  & \geq & -\ln(1-\theta). 
\end{eqnarray*}

Then, we have

\begin{eqnarray*}
\mbox{minimize} & ~ & \sum_{v \in V} x_v,\\
\mbox{subject to} &  &  x_u \geq 1, ~~~~\mbox{ for all $u \in V$ such that $deg(u)=0$},\\
& & -\ln(1-\theta) x_u ~+~ \sum_{\{v,u\} \in E} ((-\ln(P_{v,u})) \cdot x_v) ~\geq~ -\ln(1-\theta),\\ 
& & ~~~\mbox{for all $u \in V$ such that $deg(u)>0$},\\
& & x_v \in\{0,1\}, ~~~\mbox{for all $v \in V$.}
\end{eqnarray*}
where $deg(u)$ indicates the degree of node $u$.

\subsection{Probabilistic robust domination applied to the {\it C. elegans} neuronal network}
Recent reconstructions of the {\it C. elegans} neural network have significantly updated the wiring diagram of the somatic
nervous system. The new reconstruction includes original data from White et al. \cite{white}, Hall and Russel \cite{hall}
and adds new information. In particular, 3,000 synaptic contacts, including gap junctions, chemical synapses and neuromuscular
juctions were updated or added to the latest network version \cite{neuron}. As as a result, the large-scale structure
of the network has significantly changed with respect to that of White et al. Here, we focus on the connectivity
of gap junction and chemical synapse networks of {\it C. elegans} neurons. The channels that provide
electrical coupling between neurons are called gap junctions. In contrast, chemical synapses use neurotransmitters
to link neurons. Because these network are biologically different, they are treated independently, as done in \cite{neuron}.
Although it might be possible that gap junctions could conduct current in only one direction, this feature has not been observed or confirmed
yet in {\it C. elegans} \cite{neuron}. Therefore, this network was considered as undirected network. The chemical synapses, in contrast,
contains directionality capability, a feature that has been confirmed using micrographs \cite{neuron}. 
The analyzed
gap junction network consisted of 279 neurons and 514 gap junction connections. The giant connected component is composed 
of 248 neurons and two smaller components of 2 and 3 neurons. After removing the 26 isolated neurons, we performed
our analysis using 253 neurons and 514 connections. The statistical analysis revealed a power-law distribution
for the degree distribution with a characteristic degree exponent of $\gamma$=3.14 \cite{neuron}. 
The chemical synapse network consisted of 279 neurons and 2,194 directed connections. The statistical analysis showed that the in-degree (out-degree)
distribution followed a power-law with degree exponent $\gamma_{in}$=3.17 ($\gamma_{out}$=4.22), respectively. These results contrast
with analyses done using the dataset from  White et al. \cite{white}, which reported an exponential decay 
for the degree distribution \cite{amaral}.

Experimental analyses on neural networks 
have confirmed the unreliability of central synaptic transmission in rat brains \cite{failure}. The mean transmission 
failure probability was found to be $p$=0.71, with a range from 0.3 to 0.95 ($w$=0.24). In this work, we used the most well-studied
neural network corresponding to the {\it C. elegans} (chemical synapse and gap junction) to investigate probabilistic robust control.
A visual representation of experimental neural gap junction (undirected) for {\it C. elegans} 
is shown in Fig. 19. A transmission failure probability distribution similar to that observed in rat brains 
was mapped on the links of these 
networks, making a fraction of them unreliable. The results of the analyses are described in 
Figs. 16-17 for computer simulations and Figs. 20-22 for real neural gap junction and chemical 
synapse networks and suggest that the presence of variance of the failure
probability $w$ does not significantly affect the fraction of driver nodes. In contrast, it is strongly affected by 
both the minimum degree $D$ and the average failure probability $p$. This biological example of unreliable links suggests that
theoretical results and simulations on probabilistic robust control 
analysis may have an impact on understanding and controlling at will real-world systems 
with unreliable components. 

\section{Conclusion}
We have introduced the concept of structurally robust control
of complex networks and have used the MDS model, which is widely applied in engineering problems, 
to illustrate an example of robust complex network controllability. Counterintuitively, the developed analytical tools, computer 
simulations and real-world network analyses demonstrate that robust control in a large network does not 
change the order of required driver nodes compared to a conventional system without such robust capability. When 
using an MDS with $C$=1, $D$=1, the system can easily become 
uncontrollable if only one power or communication line fails during major natural disasters. In contrast, 
in the RMDS framework ($C$=2, $D$=2) the system remains controllable 
even under arbitrary single or multiple link failure. Therefore, both configurations
 require exactly the same order of controllers. Engineering and biological
systems could benefit from these findings.


In addition, the order of the MDS changes for $\gamma < 2$ by changing the minimum degree $D$ (e.g.,  
constructing real networks with degree $D>1$),
unveiling another tool to decrease the number of driver nodes. Because some real networks have 
unreliable links, we have extended our framework to 
probabilistic robust control (PMDS) and have successfully applied the developed analytical tools to real 
neural networks of {\it C. elegans} with unreliable
synaptic transmission. With the forthcoming comprehensive map of neural connections in the human brain \cite{brain2, brain}, the presented method
could offer new avenues to examine the brain's large-scale structure, to address synaptic reliability and to stimulate large fractions 
of the brain by interacting only with relatively few components.

The proposed concept of structurally robust control
of complex networks could also be investigated using a different algorithmic framework. As discussed above, we selected 
the MDS model because it has already found applications
in real engineering systems. However, the concept could also be mathematically formalised and implemented using, for example, 
the maximum matching model \cite{baraliu}. In this case, additional computations would be needed to investigate the order of drivers in 
an optimal robust control configuration; therefore, this analysis is left for future work. 

In addition, the presented method can also address the simultaneous failure of multiple links. The aim of the RMDS ($C$=2) framework is to construct a system 
that remains controllable even if an arbitrary link is damaged.
However, the developed analytical tools also allow us to design a system with a more robust configuration 
($C$=3) or ($C$=4) so that the network is still controllable
even in case of arbitrary failure of pair or triplet of links, respectively.  

The emerging picture for probabilistic failure or malfunction of transportation and transmission lines in real-world complex infrastructures, 
socio-technical networks and biological networks emphasizes the importance and role of the presented robust DS approach for controllability. 
The proposed framework and tools offer a new direction for understanding the linkage between controllability and robustness 
in complex networks, with implications from engineering to biological systems.

----------------------------------------------
J.C.N. was partially supported by
MEXT, Japan (Grant-in-Aid 25330351) and  T.A. was partially supported by MEXT, Japan (Grant-in-Aid 26540125). This work was 
also partially supported by research collaboration projects
by Institute for Chemical Research, Kyoto University.

\



\begin{figure}[th]
\includegraphics[angle=-90, width=18cm]{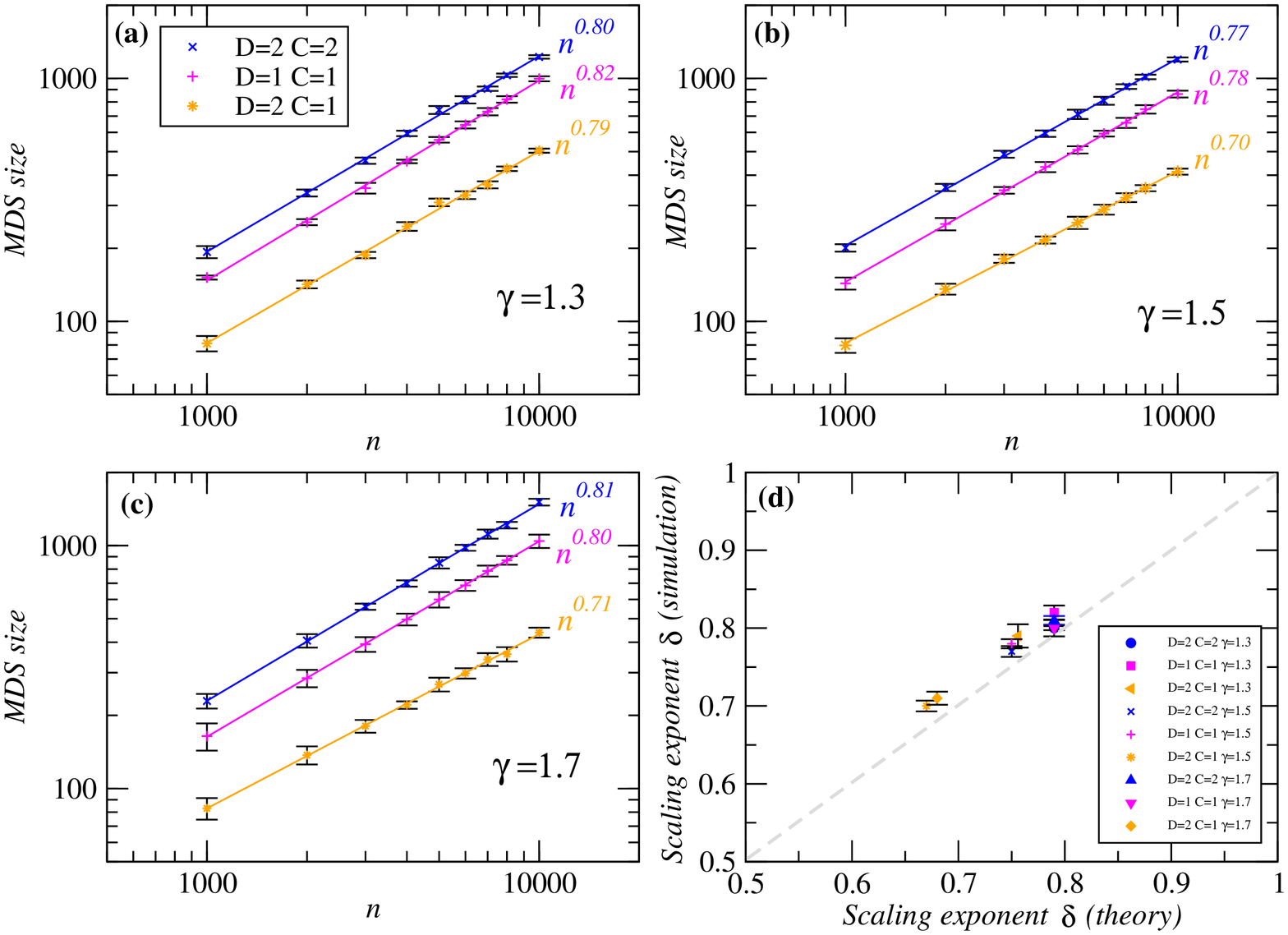}
\caption{{\bf Computer simulations for MDS size for $\gamma<2$.} The MDS size calculated in computer-generated scale-free networks 
for $\gamma$=1.3, 1.5 and 1.7 are shown in (a), (b), and (c), respectively. The configurations for 
minimum degree $D$ and cover $C$ are shown 
in the figure legend. The lines display a scaling law of $n^{\delta}$. The precise values for $\delta$ are, from top to bottom, (a)
0.804$\pm$0.011, 0.8234$\pm$0.009, 0.791$\pm$0.015, (b) 0.771$\pm$0.007, 0.781$\pm$0.006, 0.704$\pm$0.007, (c) 0.810$\pm$0.006, 0.804$\pm$0.003,
0.714$\pm$0.008. As predicted by the theory, configurations $D$=1, $C$=1 (blue) and $D$=2, $C$=2 (red) 
exhibit very similar scaling exponents. Note that $D=2$ significantly decreases the MDS size. (d) The 
scaling exponent $\delta$ predicted by theory 
compared with the scaling exponent observed in computer simulations for each $D$ and $C$ configurations.  
 The results were averaged over 10 realizations. The error bars (s.e.m.) are shown in the figure. The correlation coefficient $r$ is above
 0.999 in all cases.} 
\label{fig:combined_sim}
\end{figure}

\begin{figure}[th]
\begin{center}
\includegraphics[angle=-90, width=18cm]{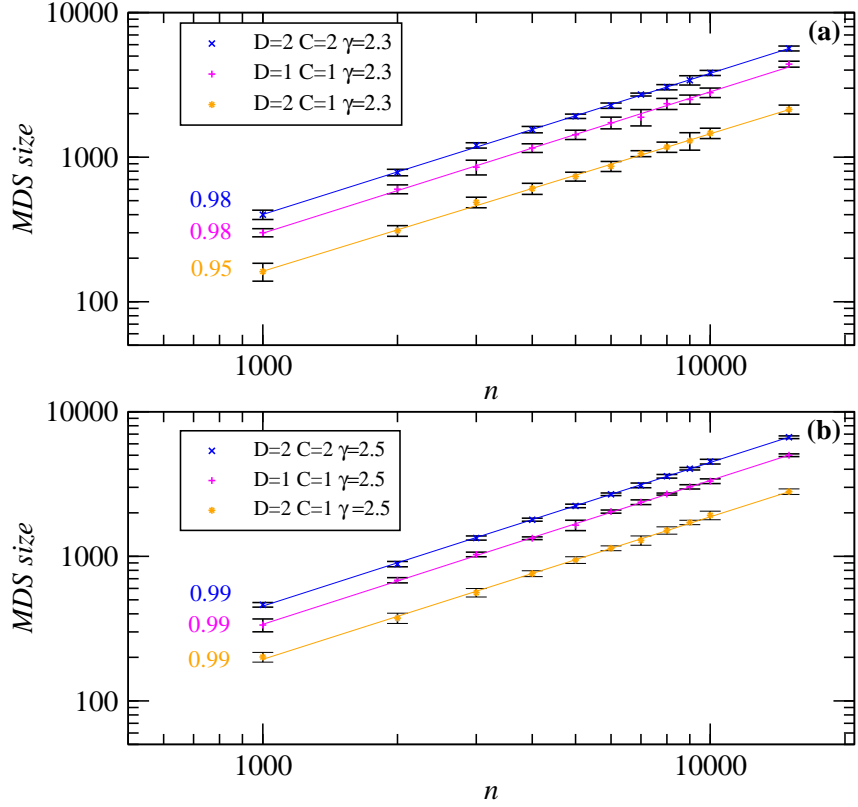}
\caption{{\bf The MDS size calculated in computer generated scale-free networks for $\gamma$=2.3, 2.5}
Configurations for minimum degree $D$ and cover $C$ are shown 
in figure legend. The lines show a scaling law as $n^{\delta}$. The precise values for $\delta$ are from up to down (a)
0.976$\pm$0.005, 0.978$\pm$0.011, 0.951$\pm$0.010, (b) 0.993$\pm$0.005, 0.995$\pm$0.005, 0.986$\pm$0.010. 
All three configurations show very similar scaling exponents close to 1, as predicted by theory.
Note that $D=2$ significantly decreases the MDS size. The error bars (s.e.m.) are shown in the figure. The correlation coefficient $r$ is above
 0.999 in all cases.}
\label{fig:23_25}
\end{center}
\end{figure}

\newpage
\begin{figure}[th]
\begin{center}
\includegraphics[angle=-90, width=18cm]{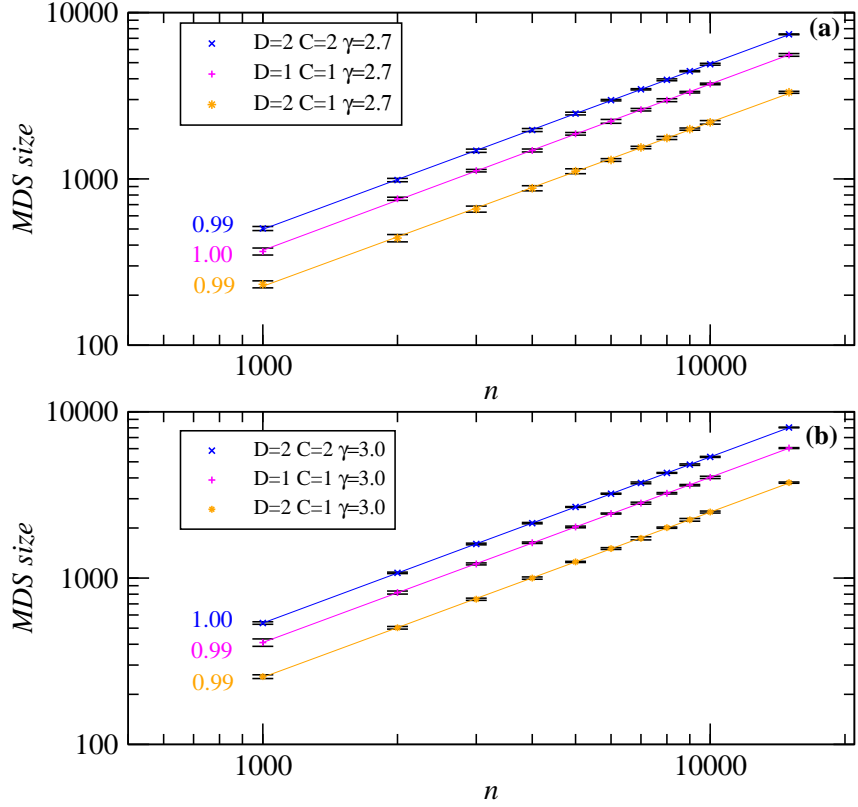}
\caption{{\bf The MDS size calculated in computer generated scale-free networks for $\gamma$=2.7, 3.0}
Configurations for minimum degree $D$ and cover $C$ are shown 
in figure legend. The lines show a scaling law as $n^{\delta}$. The precise values for $\delta$ are from up to down (a)
0.994$\pm$0.002, 0.999$\pm$0.004, 0.986$\pm$0.006, (b) 0.998$\pm$0.008, 0.992$\pm$0.011, 0.993$\pm$0.022. 
All three configurations show very similar scaling exponents close to 1, as predicted by theory.
Note that $D=2$ significantly decreases the MDS size. The error bars (s.e.m.) are shown in the figure. The correlation coefficient $r$ is above
 0.999 in all cases.}
\label{fig:27_30}
\end{center}
\end{figure}

\newpage
\begin{figure}[th]
\includegraphics[angle=-90, width=16cm]{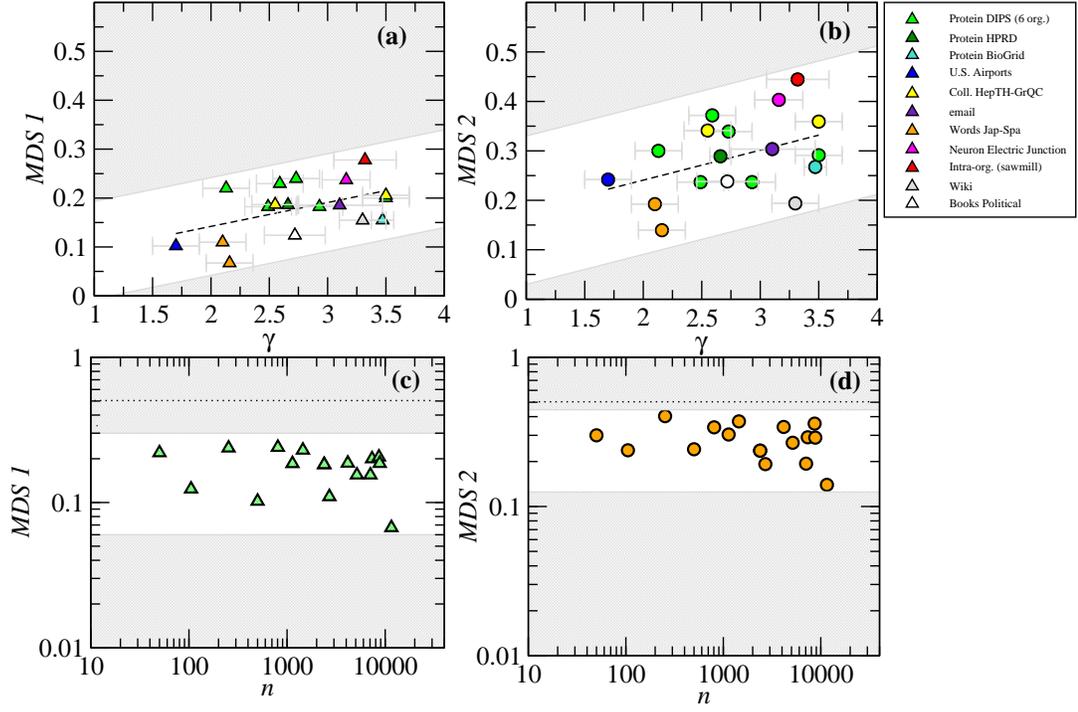}
\caption{{\bf The MDS fraction in real undirected networks.} The MDS fraction as a function of the degree exponent $\gamma$ for real undirected networks $C$=1 in (a) and $C$=2 in (b).  
The MDS fraction as a function of the network size for the same real undirected networks for $C=1$ in (c) and $C=2$ in (d). 
Highlighted regions are visual guidance of the observed tendency.}
\label{fig:undirected}
\end{figure}

\begin{figure}[th]
\includegraphics[angle=-90, width=16cm]{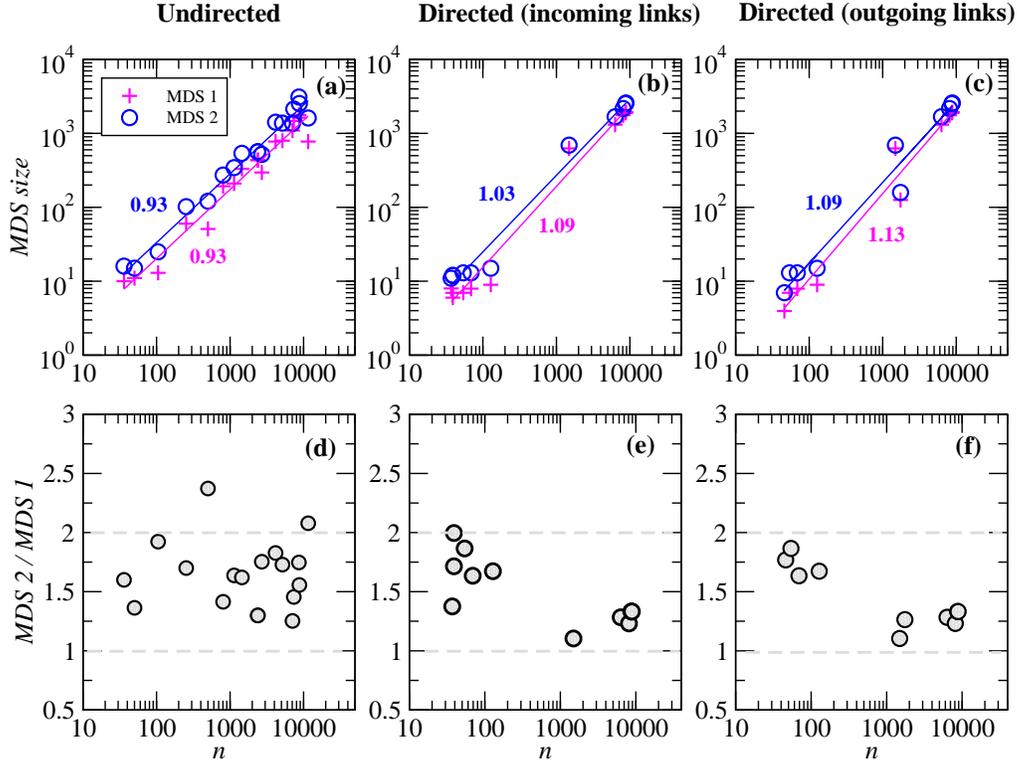}
\caption{{\bf The MDS size scaling in real undirected and directed networks.} The MDS size as a function of the network 
size $n$ for real (a) undirected and directed networks with 
(b) incoming and (c) outgoing links. The results display a scaling function of $n^{\delta}$ with $\delta$ close to 1, as predicted by 
the theoretical analysis. The precise values of the scaling exponent $\delta$ for $C$=1 (plus) and $C$=2 (circle) are (a) 0.934$\pm$0.046 
($r$=0.981), 0.931$\pm$0.036 ($r$=0.989), (b) 1.094$\pm$0.053 ($r$=0.990), 1.034$\pm$0.044 ($r$=0.992), (c) 1.134$\pm$0.071
($r$=0.984), 1.092$\pm$0.067 ($r$=0.985), respectively. The correlation coefficient $r$
is indicated between parentheses. The ratio between the MDS sizes (MDS2/MDS1) computed with covers $C$=2 and $C$=1 is shown in parts
(d-f) for the same real networks. The results show that the ratio is almost always lower than 2. See 
Tables 1 and 2 for real data details.}
\label{fig:2x3}
\end{figure}

\newpage
\begin{figure}[th]
\includegraphics[angle=-90, width=16cm]{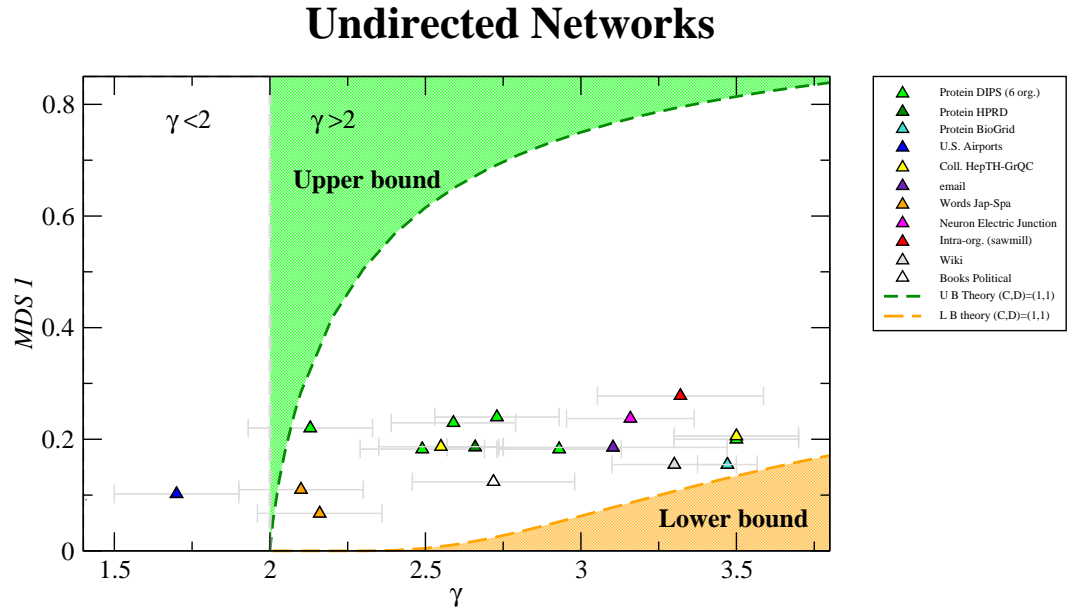}
\caption{{\bf Comparison between theoretical predicted bounds and MDS in real undirected networks.} The overlap between 
the MDS fraction with $C=1$ (MDS1) for undirected networks and the predictions
of Eqs. 6-7 for the lower and upper bounds of networks with $\gamma>2$. The real data are always within the theoretical
boundaries.}
\label{fig:theory_bounds_mapping}
\end{figure}

\newpage
\begin{figure}[th]
\includegraphics[angle=-90, width=16cm]{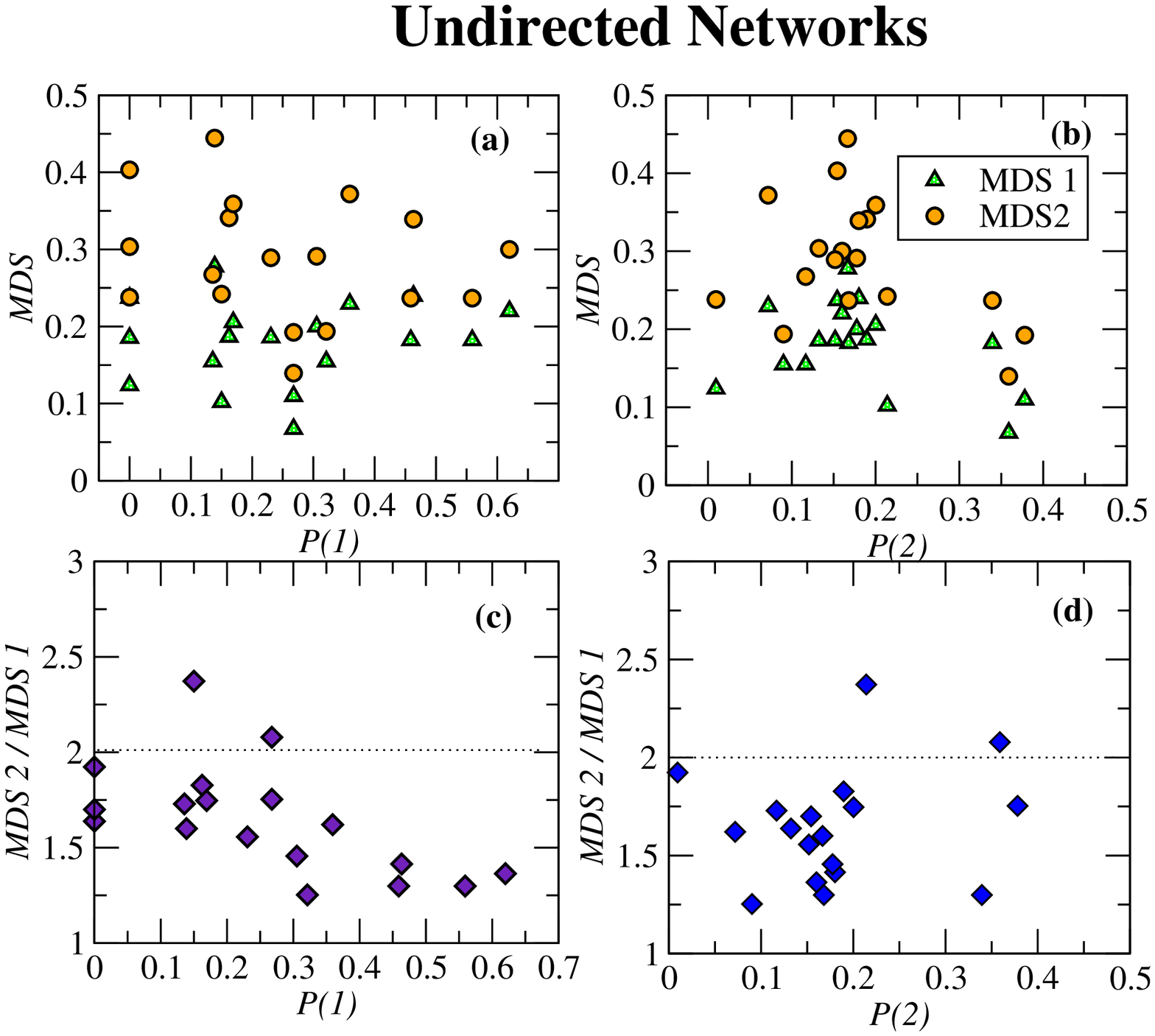}
\caption{{\bf The dependence of the MDS fraction on the degree distribution.} The MDS fraction with $C=1$ (MDS1) (a) and $C=2$ (MDS2) 
(b) as a function of the degree probability with degree 1 ($P(1)=n(1)/n$) 
and 2 ($P(2)=n(2)/n$). The ratio of MDS2/MDS1 versus (c) $P(1)$ and (d) $P(2)$, respectively. The ratio 
is below 2 in most cases.
The results demonstrate that small $P(1)$ and large $P(2)$ tend to be associated with a small MDS fraction.}
\label{fig:undirected_pk1_pk2}
\end{figure}

\begin{figure}[th]
\begin{center}
\includegraphics[angle=-90, width=16cm]{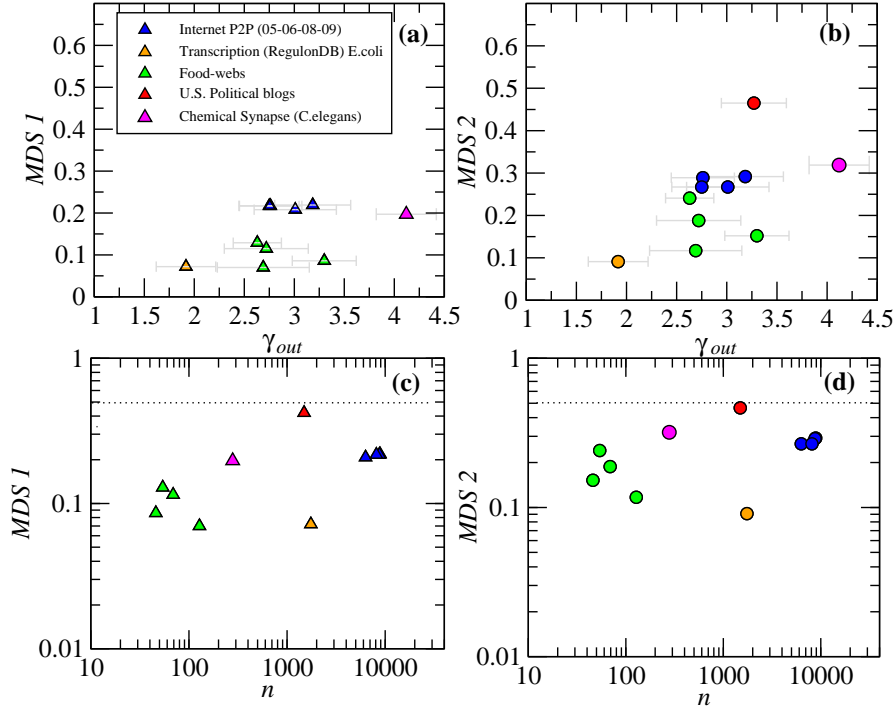}
\caption{{\bf The MDS fraction as a function of the degree exponent $\gamma_{out}$ for real directed networks $C$=1 in (a) and $C$=2 in (b)}.  
The MDS fraction as a function of the network size for the same real directed networks for $C=1$ in (c) and $C=2$ in (d). In all cases
the MDS fraction is lower than 0.5 (dotted line). Note that four out of seven food webs show power-law behavior for outgoing degrees. Each color
in circles corresponds to a network type as shown in legend. }
\label{fig:sidirout}
\end{center}
\end{figure}

\newpage
\begin{figure}[th]
\begin{center}
\includegraphics[angle=-90, width=16cm]{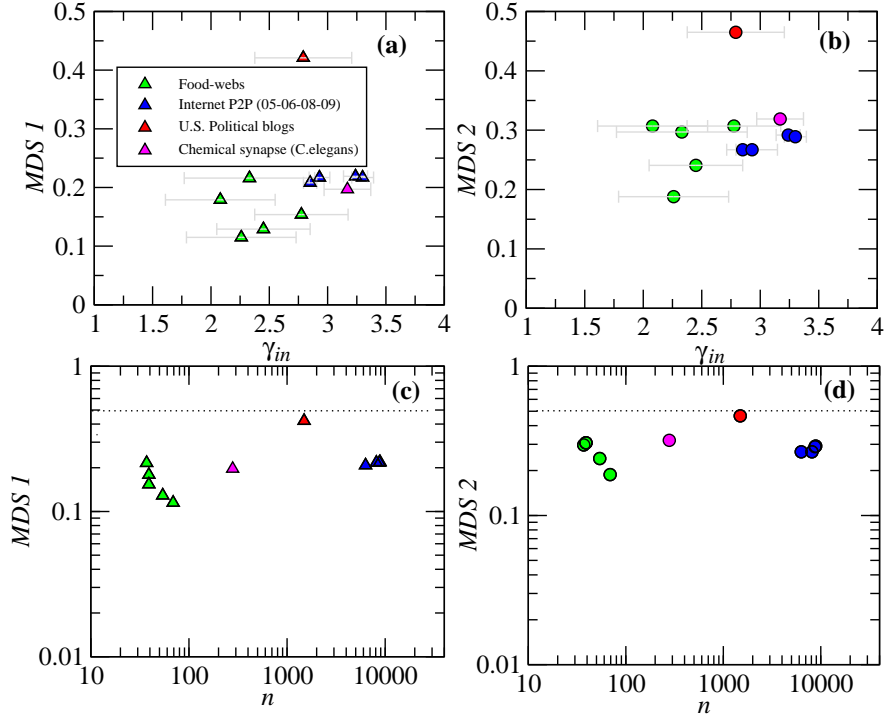}
\caption{{\bf The MDS fraction as a function of 
the degree exponent $\gamma_{in}$ for real directed networks $C$=1 in (a) and $C$=2 in (b)}.  
The MDS fraction as a function of the network size for the same real directed networks for $C=1$ in (c) and $C=2$ in (d). In all cases
the MDS fraction is lower than 0.5 (dotted line). Note that five out of seven food webs show power-law behavior for incoming degrees. 
The data for the transcriptional regulatory network is absent because it does not follow a power-law for incoming degrees.
Each color in circles corresponds to a network type as shown in legend.} 
\label{fig:sidirin}
\end{center}
\end{figure}





\newpage
\begin{figure}[th]
\begin{center}
\includegraphics[angle=0, width=16cm]{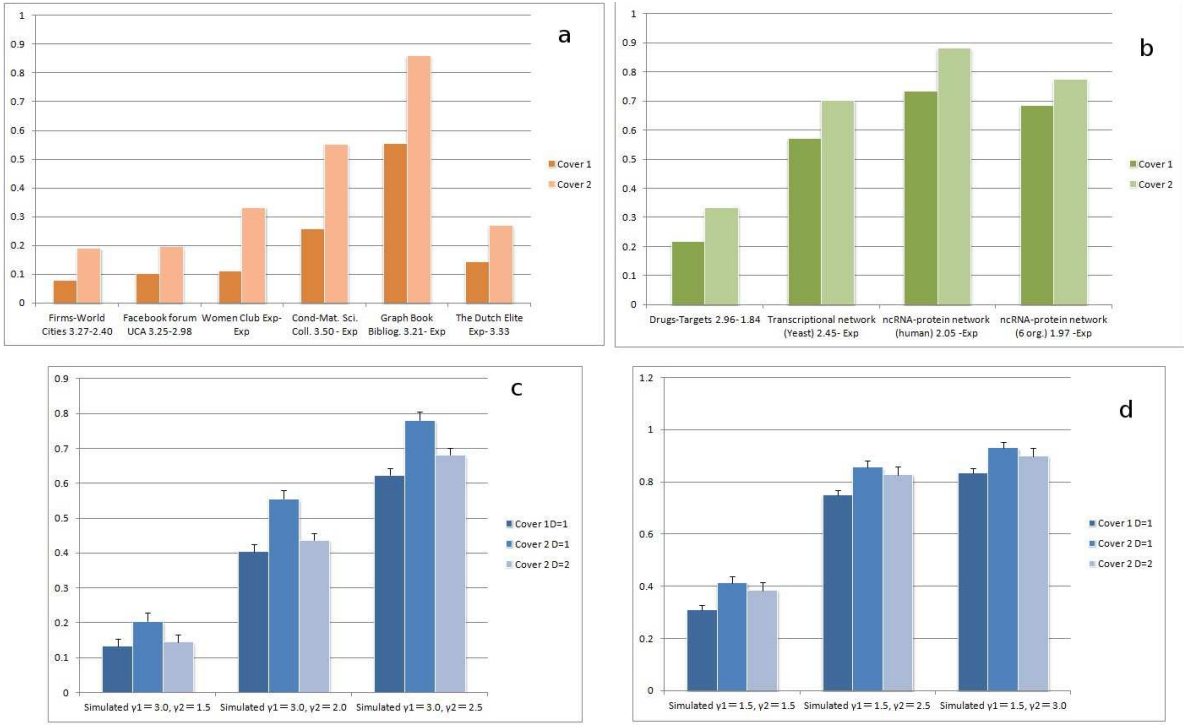}
\caption{{\bf The MDS size computed on real-world bipartite networks.} (a) Social networks and (b) 
biological networks computed for $C$=1 and $C$=2 as shown in legend. Note that social networks tend to be controlled with a smaller
fraction of nodes compared to biological networks. (c-d) Computation of the MDS in synthetic bipartite 
networks
generated with the model shown in Section IV.C with several values of degree exponents $\gamma_1$ and $\gamma_2$ and three configurations
($C$=1, $D$=1), ($C$=2, $D$=1) and ($C$=2, $D$=2). Note that the minimum degree $D=2$ almost completely compensates the increasing of the MDS size 
from the robust control $C$=2.}  
\label{fig:sibipartite}
\end{center}
\end{figure}

\begin{figure}[th]
\begin{center}
\includegraphics[width=14cm]{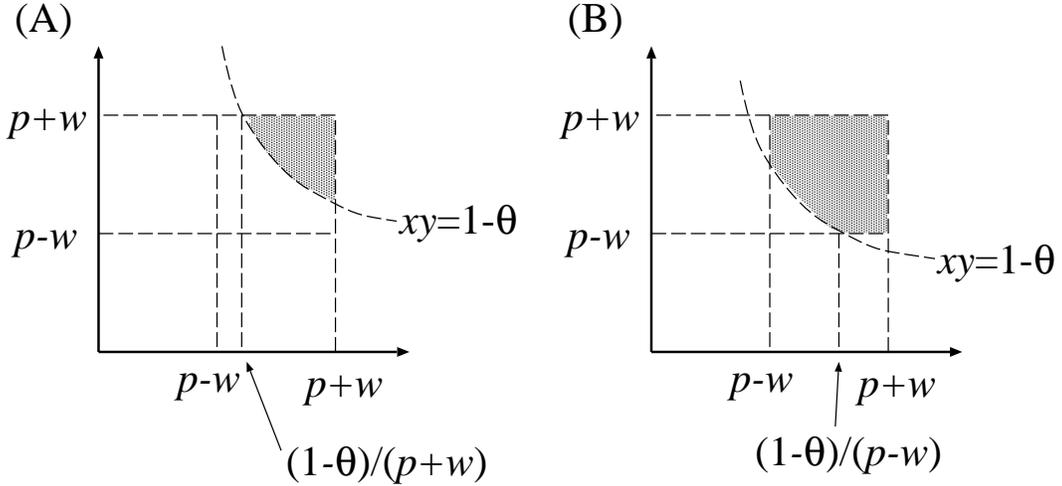}
\caption{{\bf Graphical representation of the case $D$=2 and $w>0$.} In order to estimate the fraction of degree-2 nodes to be added to a DS,
it is enough to consider the regions (A) and (B)
for the cases of $(p+w)(p-w) < 1 - \theta < (p+w)*(p+w)$ and
$(p-w)(p-w) < 1 - \theta < (p+w)*(p-w)$, respectively.}
\label{fig:deg2wplus}
\end{center}
\end{figure}

\newpage
\begin{figure}[th]
\begin{center}
\includegraphics[angle=-90, width=16cm]{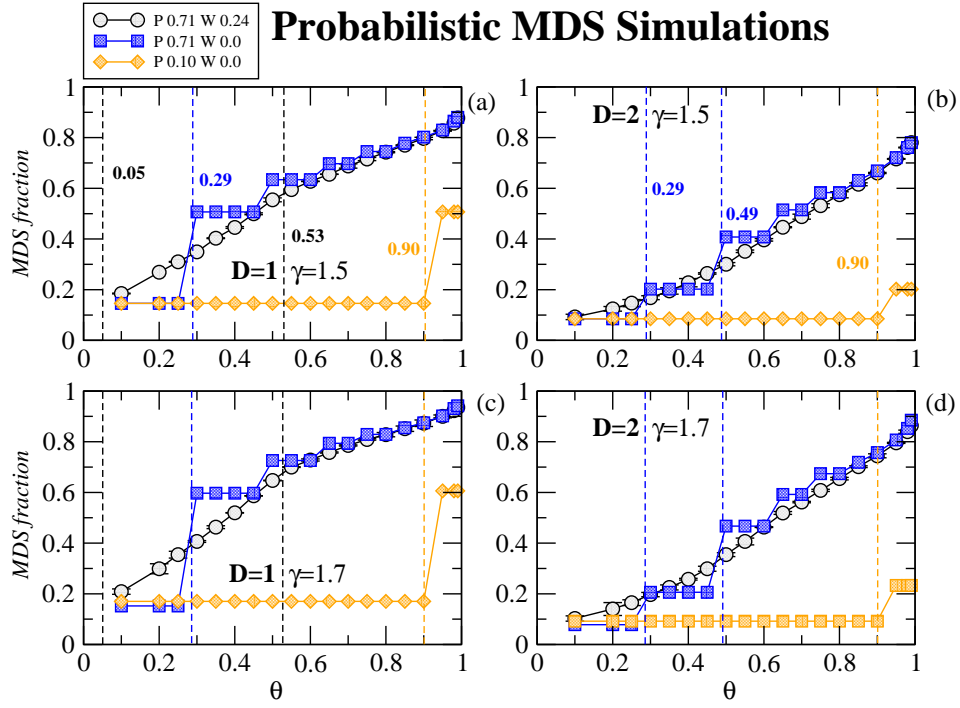}
\caption{{\bf Probabilistic MDS simulation results.} The probabilistic MDS fraction in simulated 
scale-free networks with 
($a$) $D$=1 and ($b$) $D$=2 for
 $\gamma=1.5$ and $D$=2 ($c$) $D$=1 and (d) $D$=2 for $\gamma=1.7$. The predicted theoretical thresholds (dashed lines)
 that significantly changes the MDS size are in fair agreement with observed results in computer simulations. The configurations
 for the probability of link failure $P$ and the variability change $w$ of the failure probability [$p-w$, $p+w$] 
 are shown in figure legend.}
\label{siprob1}
\end{center}
\end{figure}

\newpage
\begin{figure}[th]
\includegraphics[angle=-90, width=16cm]{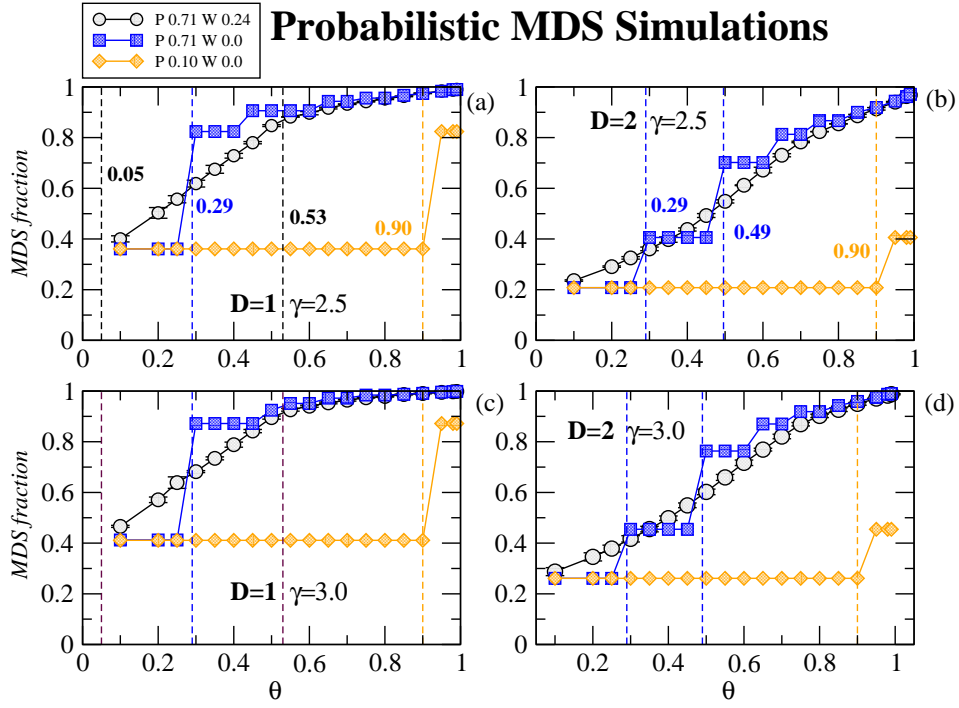}
\caption{{\bf The probabilistic robust domination on synthetic scale-free networks.} The probabilistic MDS fraction in simulated scale-free networks with ($a$) $D$=1 and (b) $D$=2 for
 $\gamma=2.5$ and ($c$) $D$=1 and (d) $D$=2 for $\gamma=3.0$. The predicted theoretical thresholds (dashed lines)
 that significantly change the MDS size are in fair agreement with the observed results from the computer simulations. The configurations
 for the probability of link failure $p$ and the variability change $w$ of the failure probability [$p-w$, $p+w$] 
 are shown in the figure legend.}
\label{fig:neuron_p71_sim}
\end{figure}

\newpage
\begin{figure}[th]
\begin{center}
\includegraphics[angle=-90, width=16cm]{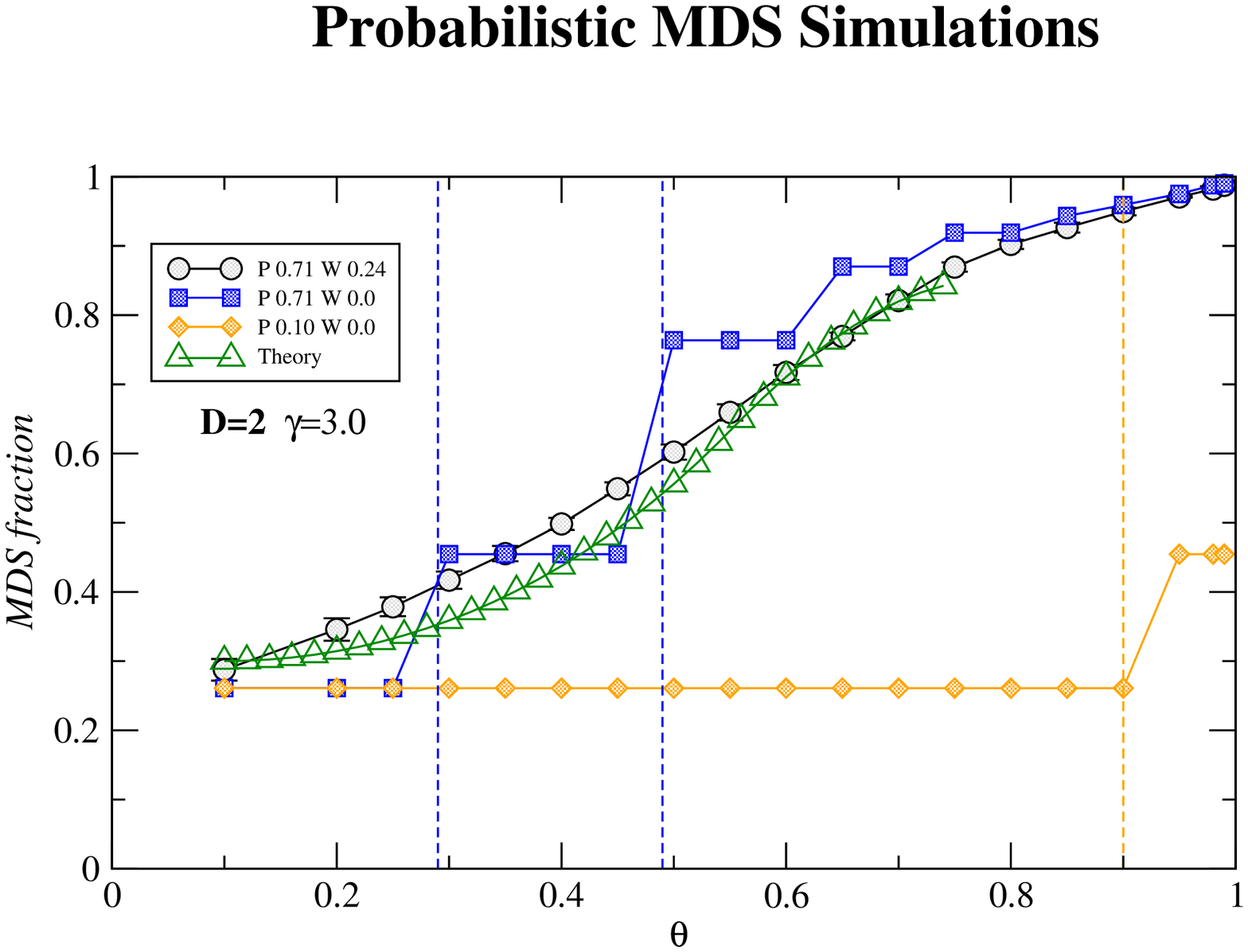}
\caption{{\bf Theoretical predictions and simulations of PMDS.} Comparison of the probabilistic MDS size computed in simulated scale-free networks
with $D$=2 and $\gamma=3$ and the case of $D$=2 predicted by theory. Theoretical values are scaled so that these take
 almost the same values as the simulated ones
 at the beginning and ending points because,
 in the theoretical analysis, it is assumed that
 all nodes are of degree 2 and
 the effects of the other nodes are ignored.}
\label{fig:siakutsu}
\end{center}
\end{figure}


\newpage
\begin{figure}[th]
\begin{center}
\includegraphics[angle=0, width=20cm]{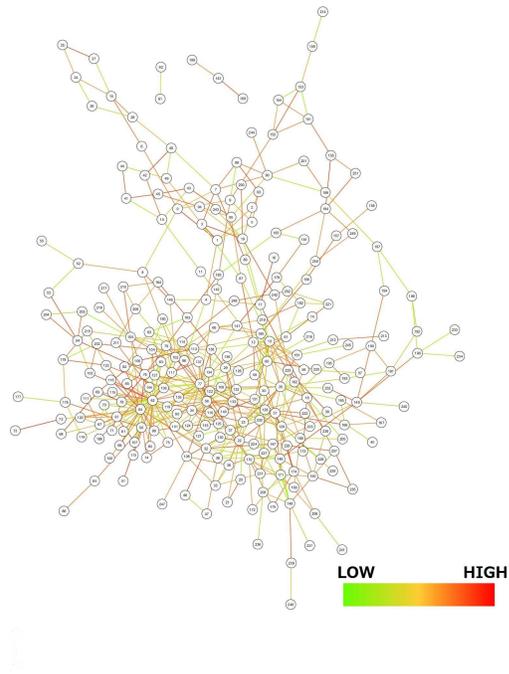}
\caption{{\bf The neural gap junction network.} Visualization of the experimental gap junction undirected network for {\it C. elegans}. 
A probability distribution of synapse transmission failure
with a peak at $p=0.71$ and width of $w=0.24$ (0.47-0.95) is mapped onto the links. }
\label{fig:sinet}
\end{center}
\end{figure}
\clearpage
\newpage
\begin{figure}[th]
\begin{center}
\includegraphics[angle=-90, width=16cm]{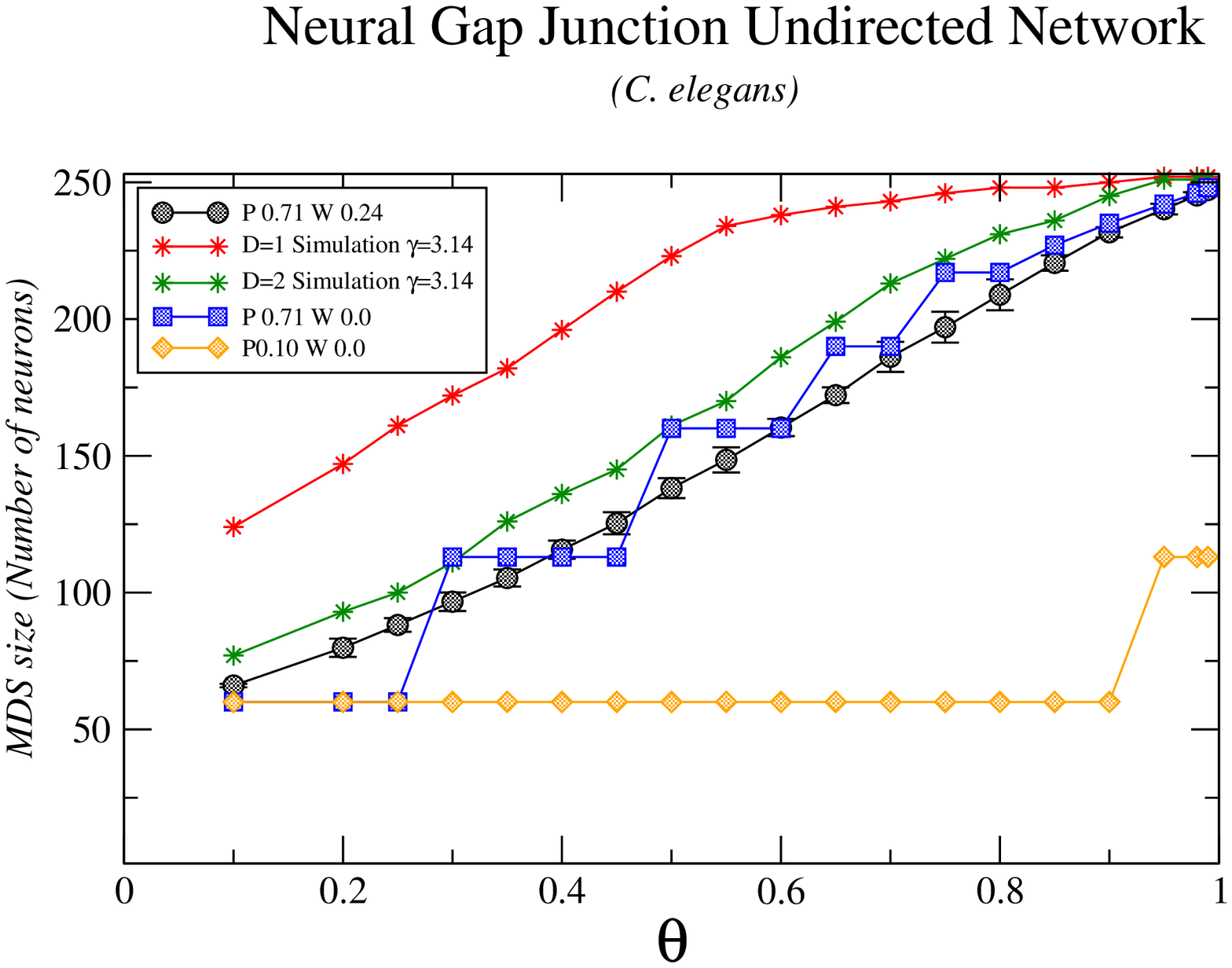}
\caption{{\bf The probabilistic MDS size computed on the gap junction undirected network.}
The computation of the probabilistic MDS size using the real neural gap junction network for $C. elegans$ organism
where a distribution of link failure was mapped on the network as shown in legend for three values of $w$ (circle, square and diamond symbols). The 
results of 
synthetic scale-free networks 
constructed using the model shown in Section III.D
and calculated with the same degree exponent and the number of nodes observed in real {\it C. elegans} network (star symbols). When simulated networks have a minimum degree $D$=2, 
the MDS size decreases as predicted by theory. Because in this simulation the synthetic network have different average degree, the
results tend to be higher than those from real network.}
\label{fig:siakutsu2}
\end{center}
\end{figure}


\clearpage
\newpage
\begin{figure}[th]
\includegraphics[angle=-90, width=16cm]{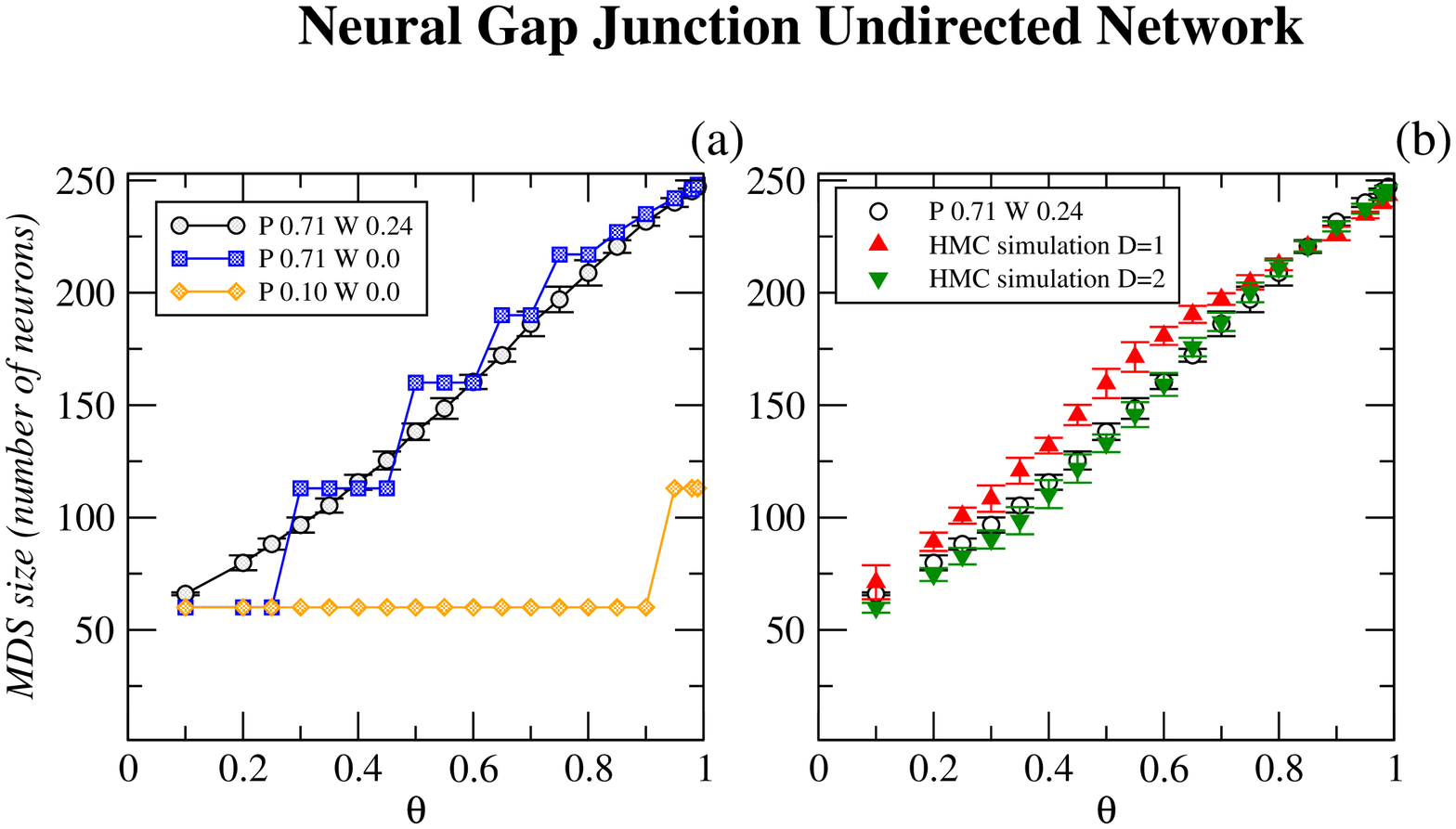}
\caption{{\bf The probabilistic robust domination on a real neural gap junction network}. The computation of the 
probabilistic MDS size using (a) the real neural gap junction network for the $C. elegans$ organism
where the distribution of link failures was mapped to the network as shown in the legend. (b) The results of synthetic scale-free networks 
constructed using the HMC model
with the same number of nodes and average degree display a similar tendency. The MDS size decreases as predicted
by theory when simulated networks have a minimum degree $D$=2.}
\label{fig:neuron_p71}
\end{figure}
\clearpage
\newpage
\begin{figure}[th]
\begin{center}
\includegraphics[angle=-90, width=16cm]{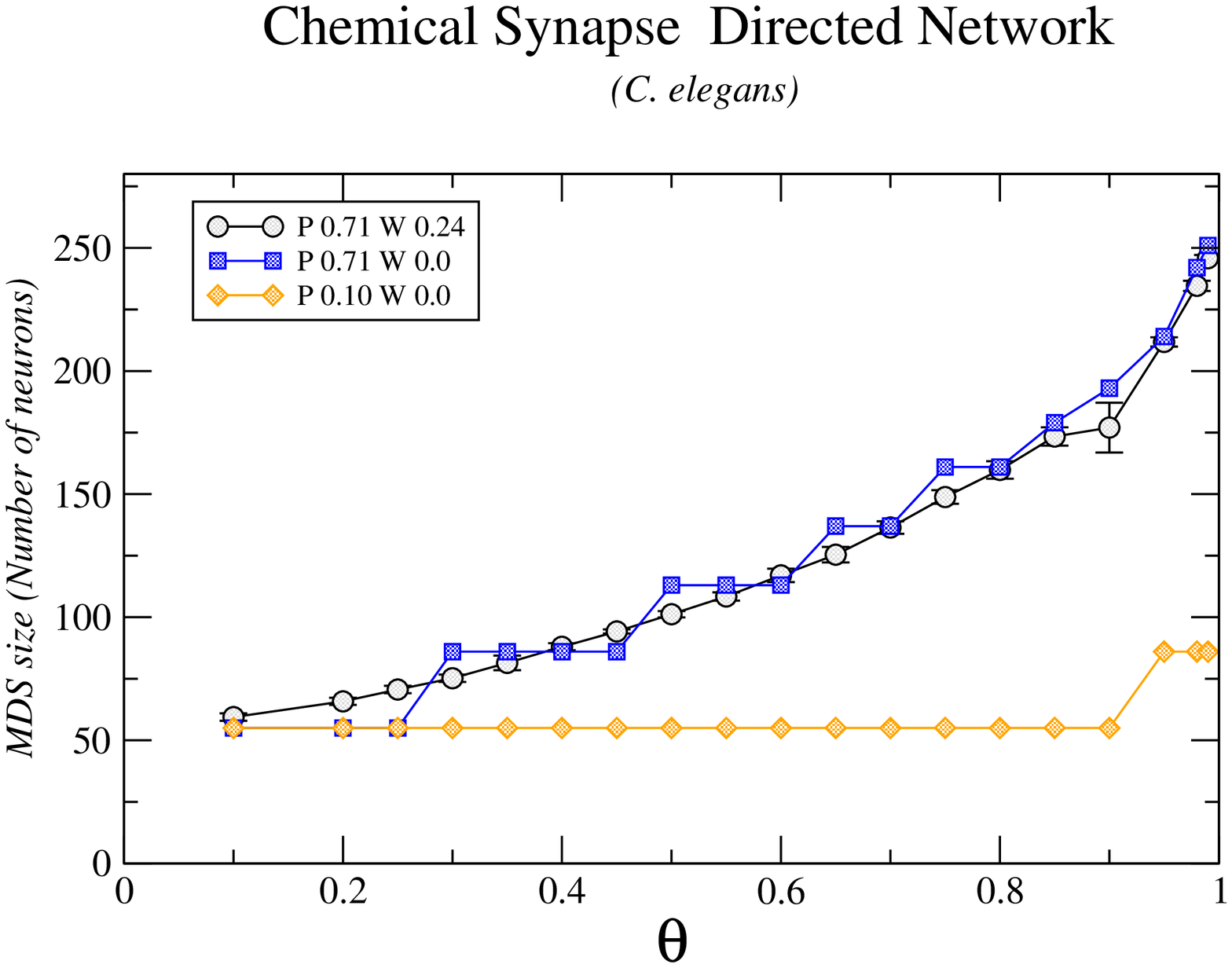}
\caption{{\bf The probabilistic MDS size computed on the chemical synapse directed network.} The computation 
of the probabilistic MDS size using the real chemical synapse directed network for $C. elegans$ organism
where a distribution of link failure was mapped on the network as shown in legend for three values of $w$. Note that the distribution follows
a curve function rather than a straight line as shown in the undirected case.}
\label{fig:sichemical}
\end{center}
\end{figure}

\clearpage
\begin{table}
\caption{{\bf The real undirected networks analysed in this work. } Type, name and description of each undirected network.
We used the discrete maximum-likelihood fitting method to estimate the degree exponent $\gamma$ from the cumulative 
degree distribution of each network \cite{m1, m2}. The standard error of $\gamma$ is derived from the width of the maximum likelihood. 
The same method was used to estimate the degree exponent 
in the directed and bipartite networks shown in Tables 2 and 3.}
\begin{ruledtabular}
\begin{tabular}{llll}
& Type & Name & Description  \\
\hline
& {\color{green} Protein}  & PPI network DIPS (6 org.) \cite{dips} & Protein networks for 6 organisms from DIPS.  \\
&  &  PPI Human HPRD \cite{HPRD} & Protein network of {\it H. sapiens} from HPRD.   \\ 
&  & PPI Yeast BioGrid \cite{BioGrid} & Protein network of {\it S. cerevisiae} from BioGrid.  \\ \hline
& {\color{blue} Transportation} & U.S. airports \cite{airport} & The largest U.S. airports connected by flights. \\ \hline
& {\color{yellow} Collaboration} & Hep-Th \cite{col} & The High Energy Physics-Theory collaboration.  \\ 
&  & Gr-QC \cite{col} & The Quantum Cosmology research collaboration. \\ \hline
& {\color{cyan} Communication} & Email \cite{email} & Email network in a university. \\ \hline
& {\color{orange} Languages} & Japanese \cite{language} & The connectivity of words in Japanese. \\ 
&           & Spanish \cite{language} & The connectivity of words in Spanish. \\ \hline
& {\color{magenta} Neuronal} & Neuronal junction \cite{neuron} &  The electric junction network of {\it C. elegans}. \\ \hline
& {\color{red} Intra-org.} & Sawmill \cite{sawmill} & A communication network within a small enterprise. \\ \hline
& {\color{light-gray} Information} & Wiki \cite{wiki} & Linked information. \\ \hline
& {\color{black} Recommendation} & U.S. politics books \cite{books} & U.S. politics books co-purchased by the same buyers. \\  
\end{tabular}
\end{ruledtabular}
\end{table}

\newpage

\begin{table}
\caption{{\bf The real directed networks analysed in this work.} Type, name and description of each directed network. The
networks whose degree distribution follows a power-law for indegree or outdegree are indicated by I or O, respectively. }
\begin{ruledtabular}
\begin{tabular}{llll}
& Type & Name & Description  \\
\hline
& {\color{blue} Internet}  & Internet P2P \cite{p2p} & Gnutella peer to peer network from August 5, 2002. \\
&   & Internet P2P \cite{p2p} &  Gnutella peer to peer network from August 6, 2002. \\
&  & Internet P2P \cite{p2p} & Gnutella peer to peer network from August 8, 2002. \\
&   & Internet P2P \cite{p2p} & Gnutella peer to peer network from August 9, 2002. \\ \hline
& {\color{orange} Gene Regulation} & Transcriptional network (O) \cite{regulon} & Transcription regulatory network of {\it E. coli}. \\ \hline
& {\color{green}  Food Web} & Cheslower (I) \cite{food} & Lower Chesapeake Bay in Summer food web.\\ 
&                  & Chespeake (I) \cite{food} & Chesapeake Bay Mesohaline food web. \\
& & Everglade \cite{food}  & Everglades Graminoid Marshes food web.   \\
&  & Florida (O) \cite{food} & Florida Bay Trophic food web. \\
& & Michigan (I) \cite{food}  & Lake Michigan food web.  \\
& & St. Marks \cite{food}  & St. Marks River (Florida) flow network.   \\
&  & Mondego (O) \cite{food}  & Mondego Estuary - Zostrea site.    \\ \hline
& {\color{red} Political} & Political blogs \cite{political} & Blog network related to politics. \\ \hline
& {\color{magenta} Neuronal}  & Chemical Synapse \cite{neuron} & The chemical synapse network of {\it C. elegans}. \\ 
\end{tabular}
\end{ruledtabular}
\end{table}

\newpage

\begin{table}
\caption{{\bf The real bipartite networks analysed in this work.} Type, name and description of each bipartite network.}
\begin{ruledtabular}
\begin{tabular}{llll}
& Type & Name & Description  \\
\hline
& {\color{blue} Social}  & Firms-World Cities \cite{world} & Services of firms across cities. \\
&   & Facebook Forum UCA \cite{facebook} &  Facebook users linked to topics. \\
&  & Davis's Southern Women Club \cite{women} & Attendance at social events by women. \\
&   & Cond-Mat Sci. Coll. \cite{condmat} & Collaboration of scientists and papers.  \\ 
&   & Graph Book Bibliography \cite{papers} & Author-by-paper network. \\
&   & The Dutch Elite \cite{pajek} & Individuals connected to administrative bodies. \\ \hline
& {\color{green} Biological} & Drugs-Targets \cite{drug_target} & Drugs binding to protein targets. \\  
&  &  Transcriptional network (Yeast) \cite{alon} & Transcription regulatory network of {\it S. cerevisiae}. \\
&  &  ncRNA-protein network (human) \cite{pinter} & Interactions between ncRNAs and proteins in {\it  H. sapiens}. \\
&  &  ncRNA-protein (6 organisms)  \cite{pinter} &  All ncRNA-protein interactions of six organisms. \\ 
\end{tabular}
\end{ruledtabular}
\end{table}

\end{document}